\documentstyle[epsf]{ICSM}
\font\tenmsam=msam10
\def\gaeq{{\hbox{\tenmsam\symbol{"26}}}}
\def\laeq{{\hbox{\tenmsam\symbol{"2E}}}}
\input epsf.tex
\begin{document}
\bibliographystyle{prsty}
\def\kvec{{\bf k}}
\def\dm{{1\over2}}
\def\sulf{(TMTTF)$_2$X}
\def\sele{(TMTSF)$_2$X}
\def\etal{{\it et al., }}
\title{\bf The normal phase of quasi-one-dimensional organic
superconductors}

\author{C. Bourbonnais and D. J\'erome$^a$\\
\it Centre de Recherche en Physique du Solide et
D\'epartement de Physique \\
\it Universit\'e de Sherbrooke, Sherbrooke, Qu\'ebec, Canada J1K 2R1.
\\ 
\it $^a$ Laboratoire de Physique des Solides, associ\'e au CNRS, B\^atiment 510
\\
\it Universit\'e de Paris-sud, 91405 Orsay, France}
\abstract{We review the properties of  quasi-one-dimensional organic superconductors:
the Bechgaard salts and their sulfur analogs  in their  normal phase  precursor to
long-range order. We  go through the main observations made in the  normal state of these systems
at low magnetic field  and tackle the issue of their description under the angles of the  
  Fermi and  Luttinger liquid pictures. }
\maketitle

\section{Introduction}

	The quest for superconductivity at always higher temperature has been a
constant motivation in the material science research over the past
quarter of the century. A well known example is the huge amount of work
which has been invested in the synthesis, measurements and processing of
intermetallic compounds belonging to the A-15 family of type II
superconducontors. These compounds V$_3$Si \cite{Hardy53} and Nb$_3$Sn \cite{Matthias54} are
still at present the materials mostly used when superconductivity comes
to applications \cite{Hulm80}. 

	In the meantime some new routes towards high temperature and possibly
exotic  superconductivity have been investigated.
This is the case for the heavy fermion compounds in which the close
interplay between local magnetic moments and the spin of delocalized
electrons has led to the possibility of a non-phonon mediated mechanism
for electron pairing \cite{Steglich79}. A very successful route towards high-$T_c$'s 
has been followed with conducting layered cuprates after the discovery of
superconductivity in (La, Sr)$_2$CuO$_4$ \cite{Bednorz86}.

	More exotic has been the proposal made by Little for the possibility of
an excitonic pairing mechanism in some organic conductors leading to a
supposedly tremendously large increase of $T_c$ \cite{Little64}. Little's
mechanism relies on the possibility to obtain an attractive interaction
between the mates of a Cooper pair in an energy range extending up to the
Fermi energy (i.e. 10,000 K) instead of the limited range imposed by the
phonon energy (100-300 K) in the well-known BCS mechanism of conventional superconductivity
\cite{Bardeen57}. A prerequisite to Little's mechanism was the synthesis of a
conducting molecular backbone with highly polarizable branched molecules.
Unfortunately $-$ or possibly fortunately $-$ a lot of physics problems had
been overlooked in the suggestion of W. Little. Molecular conduction
requires the existence of chemically stable open shell molecules able to
provide charges likely to delocalize via intermolecular overlap. A first
attempt has been reported in 1954 in perylene oxidized with bromine
\cite{Akamatsu54}. However, this conducting salt failed to present the
desired stability. Two tracks have been followed for the synthesis of
stable organic conductors. The first successful try has been obtained by
forming a compound with molecules of two different chemical nature: one
molecule that can be easily oxidized in the presence of another one
which is reduced giving rise to a charge transfer complex. A prototype
for such a system is the charge transfer compound
tetrathiafulvalene-tetracyanoquinodimethane (TTF-TCNQ), that contains
both anion (TCNQ) and cation (TTF) in the ratio 1:1 \cite{Coleman73,
Ferraris73}. The planar molecules of TTF-TCNQ form segregated stacks in a
plane to plane manner and the molecular (p-type) orbitals can interact
preferentially along the stacking direction.

\begin{figure}[ htb] 
\epsfxsize0.95\hsize
\epsffile{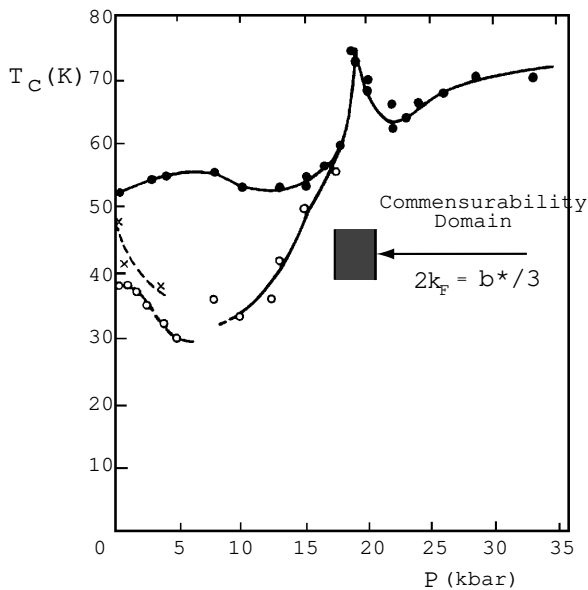}
\caption{Phase diagram of TTF-TCNQ derived by transport properties measurements, after
\cite{Friend78,Megtert79}.}    
\label{TTFTCNQ}   
\end{figure}

	In spite of the interstack interaction occurring mainly between TTF and
TCNQ stacks through S-N contacts, the dispersion of energy which derives
from the intermolecular overlap is one dimensional i.e., the dispersion is
governed only by the longitudinal wave vector giving rise to a flat Fermi
surface which is the signature of a 1-D conductor. The symmetry of the
molecular orbitals and the molecular stacking is responsible for a band
structure consisting of two inverted conduction band intersecting at the
single Fermi wave vector  $\pm k^0_F$ related to the charged $r$  transferred
from neutral TTF to neutral TCNQ by $2k^0_F= (\pi/a) r$, where $a$ is the
unit cell length along the stacking direction. The room conductivity of
TTF-TCNQ is large, $\sigma \sim  500\  (\Omega \cdot$cm)$^{-1}$
\cite{Schafer74,Groff74} and increases even more at lower temperature
reaching $10^4 \  (\Omega \cdot$cm)$^{-1}$ at 60 K $-$  although much higher
values have been claimed by Heeger and co-workers \cite{Coleman73}.
However, the dramatic increase of the conductivity breaks down at 54~K
while a transition towards an insulating ground state takes place.
Diffuse X-ray scattering experiments \cite{Denoyer75} have shown that the
54 K metal-insulator transition can be related to the instability of a 1-D
electron gas which had been proposed earlier by Peierls
\cite{Peierls55}. Attempts to suppress the Peierls state and stabilize a
conducting (and possibly superconducting state) by increasing the 3-D
character of the 1-D conductor \cite{Horovitz75} proved to be
unsuccessful since the Peierls transition is raised from 54 to 72 K under
35 kbar \cite{Jerome82}. In spite of this failure, high-pressure studies
have shown a steady increase of the charge transfer with a
commensurability regime in the 14-19 kbar pressure domain $(r = 2/3$ or
$2k^0_F = 2\pi/3a)$ (Figure \ref{TTFTCNQ}) \cite{Friend78,Megtert79}. 

However the elaboration of
a new charge-transfer conductor, TMTSF-DMTCNQ, and the use of high
pressure have provided a clue for the discovery of superconductivity. In
TMTSF-DMTCNQ, the TMTSF stacks arrange in sheets where the dominant
interchain contact is now between TMTSF stacks. Furthermore the methyl
groups in DMTCNQ molecules acting as space fillers weaken the interaction
between DMTCNQ along the stacks. Therefore this compound can be described
as an organic conductor made of only one type of stacks (the donor stack)
\cite{Jacobsen78}. The interest for this compound has been stimulated by
the high pressure investigation showing for the first time the
possibility to stabilize a highly conducting state $(\sigma \sim  10^4
\ (\Omega\cdot$ cm$)^{-1}$) at low temperature (Figure {\ref{TMDM}) \cite{Andrieux79}.
\begin{figure}[ htb] 
\epsfxsize0.95\hsize
\epsffile{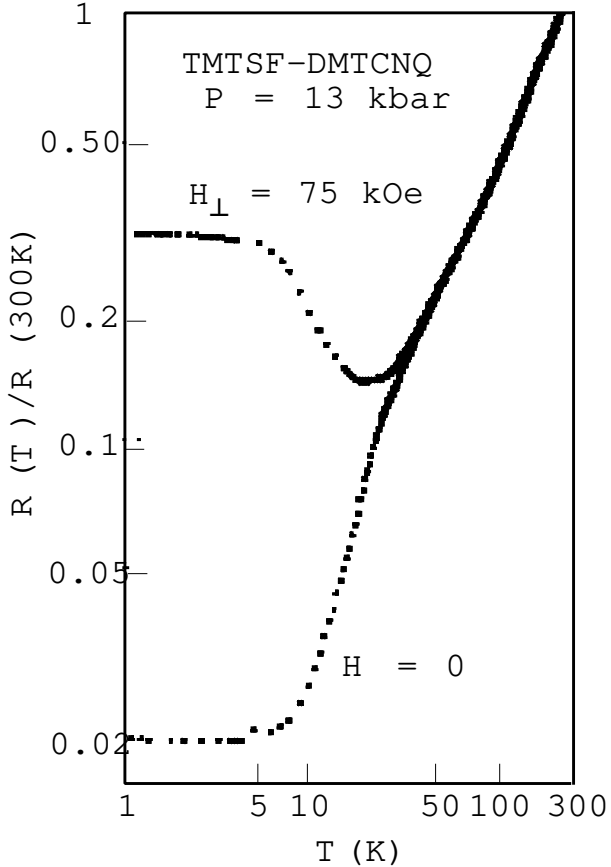}
\caption{Temperature dependence of the longitudinal resistivity of TMTSF-DMTCNQ under 13 kbar in
zero magnetic field and in a transverse field of 75 kOe, after \cite{Andrieux79}. }    
\label{TMDM}   
\end{figure}

	Given the carrier density (0.5 hole/molecule) determined from X-ray
diffuse scattering data \cite{Pouget81}, the effective one-chain
character of TMTSF-DMTCNQ and the existence of a weak $4k_F^0$ potential
coming from the packing of the DMTCNQ molecules, a new series of one-chain
organic salts (TMTSF)$_2$X, where X is an inorganic monoanion has been
synthesized \cite{Bechgaard80} $-$ this latter family, being isostructural
with the sulfur series (TMTTF)$_2$X \cite{Galigne79} (Figure \ref{Structure}).
\begin{figure}[ htb] 
\epsfxsize0.95\hsize
\epsffile{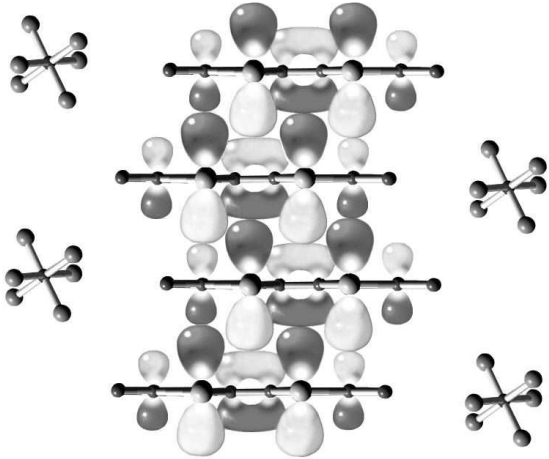}
\caption{A side view of the  (TM)$_2$X crystal structure with the electron orbitals of the
organic stack. Courtesy of J. Ch. Ricquier.}    
\label{Structure}   
\end{figure}

The possibility to vary the parameters governing the physical properties
of  (TM)$_{2}$X compounds (nature of cation or anion, pressure and
application of a magnetic field) allows an exploration  moving
continuously from half-filled band 1-D Mott insulators in sulfur  based
compounds to  conducting systems in selenium or strongly
pressurized sulfur based compounds (Figure\  \ref{Resistivity}).
\begin{figure}[ htb] 
\epsfxsize0.95\hsize
\epsffile{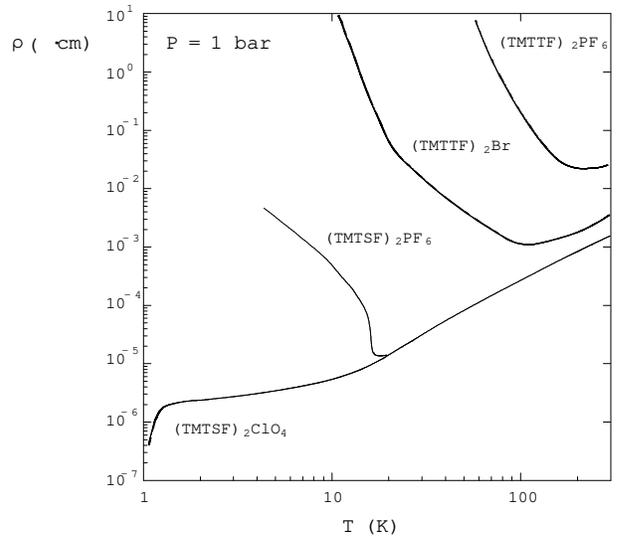}
\caption{The temperature dependence of the longitudinal resistivities for representatives of
(TM)$_2$X series. }    
\label{Resistivity}   
\end{figure} 

	One of the first compounds of the  (TM)$_{2}$X series to be synthesized
has been
(TMTTF)$_{2}$PF$_6$ \cite{Brun77}, prior to the discovery of the
Bechgaard salt (TMTSF)$_{2}$PF$_{6}$ and this salt did not raise much
enthusiasm in the early seventies because of the localized character of 
the conduction below room temperature instead of the expected metallic
behavior \cite{Coulon82}. However this compound  gained much interest in
time since it is the member in the  (TM)$_{2}$X family which spans the
broadest variety of physical properties while changing the pressure
\ \cite{Jerome91}. At low pressure, members of the sulfur series can develop either spin-Peierls or
commensurate-localized antiferromagnetic long-range order, while  either itinerant
antiferromagnetism or superconductivity are found in the selenide series. Under pressure the
properties of  the sulfur series evolves  toward those of the selenides.

 A large  number of theoretical works    have been devoted to
evaluate the importance of Coulomb interaction in the description of the
normal state of the Bechgaard salts.  At the very outset, that these
systems  show in several circumstances long range antiferromagnetic
ordering at sufficiently low temperature does indicate that repulsive
interaction among carriers is important. Although a rather large
consensus  quickly emerged  about the relevance of short-range Coulomb
interactions  in these systems, opinions differ, however,  as to the  size
of these interactions relative to the bandwidth and their role when the
temperature  is raised outside the critical domain linked to a phase
transition and where the system enters in
  normal state. The reason essentially fits in the fact that the band
structure of these systems is characterized by one-electron transfer
integrals $t_a$ along the chains which are at least one order magnitude
larger than the corresponding integrals $t_\perp$ in the transverse
directions making them close realizations of one-dimensional systems.
In  a one-dimensional metal, interactions have very special if not
dramatic consequences on the nature of the electronic state, which turns
out to be quite different  to what we are used to expect in  ordinary 
metals in three dimensions.   In  metals, like Cu, Al,
\ldots for example, the Pauli principle considerably restricts the
possibility of  scattering events in the proximity of  the Fermi sphere
so the Coulomb repulsion, though strong in amplitude, has very little
effect on low-energy carriers which behave as effective free particles,
called  `quasi-particles'. At the heart of  the celebrated {\it Fermi
liquid}  (FL) theory of the electron state, the concept of
quasi-particles proved to be inapplicable  when carriers are forced to
move and interact within a one-dimensional world. In their linear motion,
electrons can hardly avoid each other and the influence of interactions
becomes  magnified to such an extent that quasi-particles are completely
absent from the low-energy excitation spectrum. This occurs   to the
benefit of spin and charge collective modes which  form a
distinct electronic state called a {\it  Luttinger liquid} (LL) \cite{Haldane81}.
\begin{figure}[ htb] 
\epsfxsize0.95\hsize
\epsffile{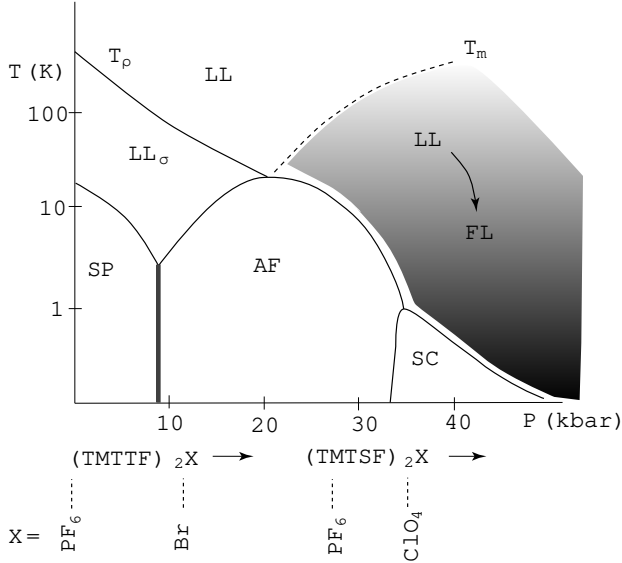}
\caption{The generic phase diagram of the (TM)$_2$X as a function of pressure or anion
substitution. On the left, the normal phase of sulfur compounds can be described as a Luttinger
liquid that  becomes gapped in the charge sector (LL$_\sigma$) below $T_\rho$ and can
develop either a spin-Peierls (SP) or localized antiferromagnetic ordered state. Under pressure,
the properties of the sulfur series evolve toward those of the selenides for which the normal
state shows a progressive restoration of a Fermi liquid (FL) precursor to antiferromagnetism (AF)
and superconductivity (S), after \cite{Bourbon98}.}    
\label{Diag-phase}   
\end{figure}

It follows that the evaluation of the extent to which one-dimensional
physics is relevant  has always played an important part in the debate
surrounding the  theoretical description of the  normal state of these
materials.   One  point of view expressed is that the amplitude of
$t_{\perp b}$ in  the $b$ direction  is large enough for  a FL component  
to develop in the ab plane, thereby  governing most properties of the 
normal phase attainable below say  room temperature. In this scenario,
the anisotropic Fermi liquid then constitutes the basic electronic
state from which various instabilities of the metallic state, like
spin-density-wave, superconductivity, etc., arise \cite{Gorkov96}. Following the
example of the BCS theory of superconductivity in conventional
superconductors, it is the critical domain of the transition that
ultimately limits the validity of the Fermi liquid picture in the low
temperature domain.

 Another  view is to consider the Bechgaard salts as more correlated
systems as a result of one-dimensional physics which maintains 
relevance within a large extent of the normal phase  and even
affects the mechanisms by  which long-range order  is stabilized in these materials.

The  literature devoted to both experimental and theoretical
aspects of (TM)$_2$X over the past eighteen years or so, is far too vast to be presented from the
top to the bottom within the present short review. A selected interest will
thus be put  on the nature of  the normal phase of the (TM)$_2$X compounds. This
issue has actually  a long history which  traces back to
the discovery of organic superconductivity and to the first attempts to figure out  the origin of
this phase in these compounds \cite{Jerome80}. It was later on tied with the more general frame of
the physics of the normal phase in  strongly correlated anisotropic superconductors including 
high-$T_c$ cuprates.   Even within  this restricted scope we will not dwell on the particular 
behavior of the normal phase taking place at   low temperature under magnetic field. We refer
the reader  to recent references  on various aspects of anomalous magnetotransport in the (TM)$_2$X
(see for example
\cite{Chaikin98,Behnia95,Clarke97,Yakovenko98,Naughton98}).

\section{Single-particle band theory}
 	The metallic
character of the organic conductors to be discussed in this article occurs
through the delocalization of unpaired carriers via the overlap of
$\pi$-orbitals between neighboring molecules (Figure \ref{Structure}). The organic salts of the
(TM)$_{2}$X family represent archetypes of 1-D conductors where TM is
usually tetramethyltetraselenafulvalene TMTSF, a
flat molecule that donates electrons easily and X is a monovalent anion,
which either consists   of symmetric (PF$_{6}$,
AsF$_6$, etc.) or asymmetric (ClO$_{4}$,
ReO$_4$, NO$_{3}$, FSO$_3$, SCN, etc.) inorganic molecules.

	What is remarkable with the (TM)$_{2}$X series is the variety of
different physical situations that can be achieved while changing the
chemical nature of the organic donor, i.e. using
tetramethylthiafulvalene TMTTF, the sulfur analog of TMTSF that
contains four sulfur atoms instead of selenium atoms, changing the nature
of the anion (its size or symmetry), synthesizing sulfur-selenium hybrid
donors or even ordered (or disordered) solid solutions of TMTTF and
TMTSF molecules \cite{Batail94}. All above mentioned compounds are
isostructural and crystallize in the triclinic
$P\bar{1}$  space group with two donors and one anion per unit cell.
Since most electronic density lies symmetrically on both sides of the flat
TM molecules, the best optimization to the cohesive energy coming from
the delocalized carriers is achieved when the donors stack is in a plane to
plane configuration giving rise to an electron delocalization along the
stacking direction.
Two organic molecules per  unit cell belonging  to the same stack are 
oxidized and
contribute one electron to each unit cell. 

Since  flat TM organic molecules   are not strongly interacting
objects in the solid state  this confers  some relevance  in the
local point of view  in the  band structure calculation. 
In the present case, the starting point is the   
 HOMO state  (highest occupied molecular orbital ) which is constructed
from a  linear combination of atomic orbitals of the TM molecule (Figure \ref{Structure}).  The use
of the extended-H\"uckel method of quantum chemistry  then leads to an   appropriate description of
the band formation in these narrow band systems.  Most band structure calculations reported so far,
however, have been made using
 a 2-D model \cite{Grant83,Yamaji82,Ducasse86}, 3D calculations have
been performed more recently for
(TMTTF)$_{2}$Br and (TMTSF)$_{2}$PF$_{6}$ \cite{Canadell94}. The
tight-binding band of all compounds looks like the typical dispersion
relation in Fig. \ref{bande}. 
The band structure parameters thus obtained can be  used  to define the 
following model of the energy spectrum:
\begin{equation}
\epsilon( \kvec)= -2t_a\cos(k_a a/2) -2t_{\perp b}\cos (k_{\perp b }b) - 2t_{\perp
c}\cos(k_{\perp c} c).
\label{spectrum}
\end{equation}
where  it  is assumed that the underlying lattice is
orthorhombic. The  symmetry of the lattice 
 in the (TM)$_2$X  being
triclinic, the above expression then represents  a  
simplified model of the actual spectrum  of Figure \ref{bande} but it
retains the essential and  is easier to manipulate.   The conduction
band along the chain direction has an overall width $4t_a$ ranging
between 0.4 and 1.2 eV, depending on the chemical nature of the donor
molecule. As the overlap between the electron clouds of neighboring
molecules along the stacking direction is about 10 times larger than the
overlap between the stacks in the transverse $b$ direction and 500 times
larger than that along the $c$ direction the electronic structure can be
viewed at first sight as one-dimensional with an open and slightly warped
Fermi surface centered at the Fermi wave vector 
$\pm k^0_F$  defined  for  isolated chains (Figure \ref{bande}). 
\begin{figure}[ htb] 
\epsfxsize0.95\hsize
\epsffile{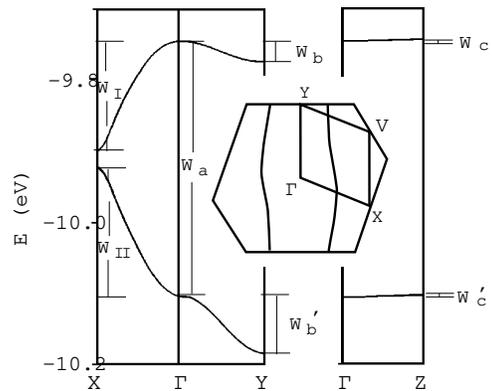}
\caption{Electronic dispersion relation and projected 2D Fermi surface  for (TMTTF)$_2$Br
calculated on the basis of its room temperature and ambient pressure structure, after
\cite{Canadell94}.   }    
\label{bande}   
\end{figure}  
\begin{table}[ htb] 
\epsfxsize0.95\hsize
\epsffile{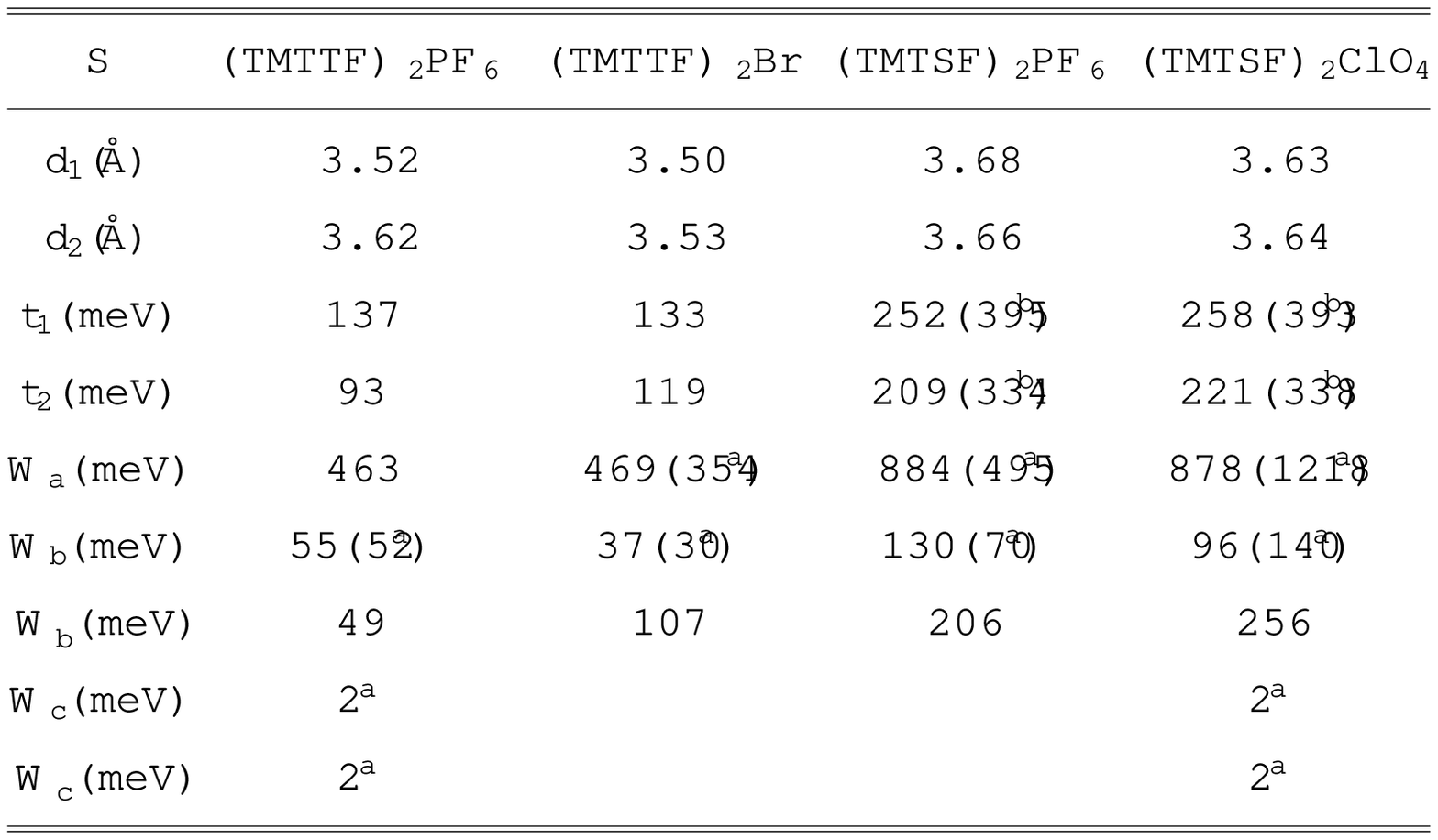}
\caption{ Intrastack crystallographic data and  calculated band parameters for members of the
(TM)$_2$X series, after \cite{Ducasse86}, $^a$\cite{Canadell94}, and  $^b$\cite{Grant83}.   }    
\label{Tableau}   
\end{table}

The anions located in centrosymmetrical cavities lie slightly above or
below the molecular planes. This structure results in a dimerization of
the intermolecular distance (overlap) with a concomitant splitting of the
HOMO conduction band into a filled lower band separated from a
half-filled upper (hole-like) band by a gap  $\Delta_D$  at $\pm 2k^0_F$,
called the dimerization gap which is shown in Fig.~\ref{bande} at the
point X of the new Brillouin zone. However, on account of the transverse
dispersion, this dimerization gap does not lead to a genuine gap in the
density of states as shown from the extended-H\"uckel band calculation
(Figure \ref{DOS}). The only claim which can be made is that these
conductors have a commensurate band filling (3/4) coming from the 2:1
stoechiometry with a tendency towards half filling which is more
pronounced for sulfur than for selenium compounds, while it  differs 
from compound to compound  
 within a given series.
\begin{figure}[ htb] 
\epsfxsize0.95\hsize
\epsffile{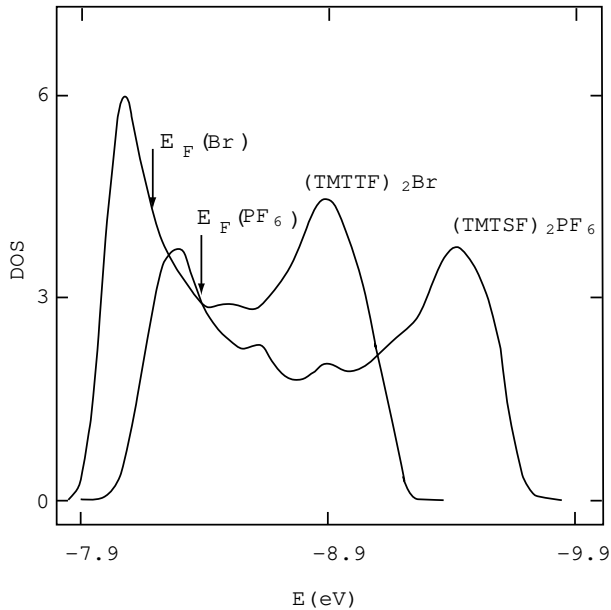}
\caption{Electronic density of states of (TMTTF)$_2$Br and (TMTSF)$_2$PF$_6$. Courtesy of E.
Canadell. }    
\label{DOS}   
\end{figure}

\subsection{Role of anions}
  The
possibility for (TM)$_2$X compounds having non-centro-symmetrical anions 
to undergo a structural phase transitions can modify the band
structure and the topology of the Fermi surface.
Anions such as ClO$_4$, ReO$_4$, NO$_3$, SCN etc., have   two
equivalent
orientations corresponding to short and  long contacts  between
Se (resp. S) atoms of TMTSF (resp. TMTTF) molecule and a peripheral
electronegative  atom of the anion.  Consider the case of  
(TMTSF)$_2$ClO$_4$, the anion
lattice orders at 24K leading to  a superstructure of the Se-O contacts
with a wave vector {\bf q}$_A$=(0,$\dm$,0) $-$ here expressed in
units of the reciprocal lattice vector \cite{Moret85,Pouget96}. The periodic
potential thus created connects two Fermi points along the $b$ direction
and   opens a gap which doubles the unit cell along that direction.
The folding of the Fermi surface that results introduces two warped
Fermi surfaces near $\pm k_F^0$. Anion lattice superstructure  has
thus important consequences on one-particle spectrum and its symmetry
properties $- $ the nesting  of the Fermi surface. This plays an
important role in the strength of electron-electron interactions at low
temperature. It also influences  the stability of the superconducting
phase  in (TMTSF)$_2$ClO$_4$ below 1.2 K at ambient pressure and the
variety of features induced by a  magnetic field such as the so-called
quantization of the Fermi surface  nesting  in the  spin-density-wave
phase ordering
\cite{Naughton88,Chaikin95,Wang93,Berthier98}, and the  Lebed
resonances \cite{Lebed96}. This  is particularly manifest when it is
compared  to compounds with spherical anions such as PF$_6$, AsF$_6$,
etc.,  for which the absence of alteration of the Fermi surface  via
anion ordering
 entails for example the stabilization of spin-density-wave long-range 
order at ambient pressure.

 For  other compounds with  a non-centrosymmetrical anion like
ReO$_4$, the  structural ordering  is different and takes place at {\bf
q}$_A$= ($\dm,
\dm,\dm)$; its impact on the electronic structure, however, turns out to
be   more marked since  the anion potential  at  this wave vector creates
a gap over the whole Fermi surface which is  so large in amplitude $(\sim
t_a$) that it leads to an insulating state in which electron-electron
interactions probably play little role.  The application of hydrostatic
pressure is then required to  restore the
metallic state and the possibility of long-range ordering for electronic
degrees of freedom \cite{Moret86}.

As previously mentioned, the anion 
potential produced by spherical anions like  
PF$_6$, AsF$_6$,..., leads to a modulation  of the charge along the
organic  stack with the same  periodicity as the
dimerization \cite{Wudl83}. It may independently contribute to the half-filled
character of the  band and then enhances the strength of
electron-electron interaction  at low temperature \cite{Emery83}.
\section{Normal state: Fermi liquid description }
Although the relevance of electron-electron interactions  is in general
not  disputed in (TM)$_2$X, it is on the point of their
magnitude and  their magnification $-$ as a consequence 
of   low-dimensionality   in a experimentally accessible part the
normal phase $-$ that  opinions  differ. 
Let us consider the point of view where   
the interacting fermions of  the normal phase are described within the 
Fermi liquid picture. By doing so in the present case, all the
complications of one-dimensional physics and the concomitant correlation
and Luttinger liquid effects turn out to be  relegated at  high-energy or
temperature.

 A rough, zero-order, estimate of the extent to which  a
Fermi liquid description would be  viable in the normal phase is
provided by
 the scale  of
$t_{\perp b}$ given by  band calculations.   Consider the temperature
range 
$T
\ll t_{\perp b}$ ($k_B=1$),  where thermal fluctuations are sufficiently
weak  to lower the uncertainty on the transverse band  wave vector
  to a range of  values $\delta k_{\perp b}  \ll 1/d_{\perp b}$, that
is small compared to the size of Brillouin zone. 
 The band wave vector $k_{\perp b}$ is therefore a good quantum number
so  the transverse band motion and the curvature of the Fermi
surface are 
 coherent. Otherwise. when
$T
> t_{\perp b}$, one has
$\delta k_{\perp b}
> d^{-1}_{\perp b}$, which is large enough for the curvature of
the Fermi surface  to be
blurred by thermal effects;   quantum coherence of electrons is then
perpendicularly disrupted   and  is thermally confined to
small distance
$\delta r_{\perp b} < d_{\perp b}$ (which actually coincides
with the perpendicular de Broglie wavelength), thereby making the
relevant  physics  essentially one-dimensional in character. In this
simple description the characteristic
temperature scale $T_{x^1}\sim t_{\perp b} $ then signals a gradual change of 
dimensionality or a crossover in the properties at the one-particle
level  \cite{Emery83}.

  According 
to band calculations (see Table 1), $t_{\perp b}$ is
around 200K  in systems like (TMTSF)$_2$X, while it is slightly less in
\sulf. This figure being large compared to the corresponding scale  for
long range order ($\sim$1K\ldots, 20K), most part of  the normal state
below room temperature  should then be  correctly described 
 by a Fermi liquid picture. In this
scenario, the quasi-particle motion, albeit anisotropic, is influenced
by the interactions with other particles through some kind of a
mean-field effect. The energy spectrum thus keeps the same form as
(\ref{spectrum}), except for a renormalization of the effective mass of
the quasi-particle in each direction. One can then write for the spectrum 
\begin{equation}
\epsilon^*_p({\bf k})-\mu = v_F^*(pk-k^0_F) - \sum_{i=b,c}2t_{\perp i}^*
\cos (k_{\perp i}d_{\perp i}),
\label{flspectre}
\end{equation}
where in the spirit of the Fermi liquid theory, the longitudinal part of
the spectrum has been linearized around the 1D Fermi points $pk^0_F =\pm
\pi/2a$,
$v_F^*=k^0_F/m^*\  (\hbar=1)$ is the effective Fermi velocity and $m^*$
the effective mass along the chain direction. As for the transverse part,
we have kept the tight-binding structure of the spectrum which is
assumed to be unaltered by the physics taking place at high energy. This
form is essential in order  to preserve the symmetry properties of the
spectrum and  a open Fermi surface.   
 Here $t_{\perp b,c}^* $ denote the effective $-$ renormalized $-$  
transfer integrals along $b$ and $c$ directions. This renormalization,
which actually results from the `history' of the system at high
energy $-$ where it is presumably 1D $-$  will in turn  affect the
crossover temperature that is, $T_{x^1}\sim t^*_{\perp b}$ \cite{Bourbon84}.

\subsection{ Fermi liquid properties at equilibrium} 

\paragraph{Susceptibility.\ $-$}
In the Fermi liquid picture, static and uniform response function like
the spin susceptibility
$\chi_s$  is also renormalized with respect to the ideal Fermi gas
prediction. It takes the form \cite{Pines66}  
\begin{equation}
\chi_s =  {2 \mu_B^2N(E^*_F)\over 1+ F^a}, 
\label{FLchi}
\end{equation}
where $\mu_B$ is the Bohr magneton and $N(E^*_F)= (\pi v_F^*)^{-1} $ is
the electronic density of states per spin. Here $F^a$ is 
an effective coupling  constant   which favors spin alignment and
then leads to an enhancement of the susceptibility for repulsive
interaction (\hbox{$F^a<0$} ).  Although this constant is 
phenomenological in the framework of the Fermi liquid theory, 
a straightforward
connection between 
$F^a$ and a microscopic model like the Hubbard
model can be made through a first-order Hartree-Fock  calculation $-$
which is  actually equivalent to a static Random Phase Approximation
(RPA). One then finds $F^a = - N(E_F) aU
$, where $U$ is the short-range Coulomb repulsion parameter of the Hubbard model. 
Thus in the Hartree-Fock picture, the   amplitude of  $U/4t_a$ for
different compounds can be extracted from the ratio $\chi_s/\chi_s^0 $
between   the observed susceptibility in the very low temperature
region
\cite{Miljak83,Miljak85} and its calculated value using the band parameters
extracted from 
experiments or from band calculations
\cite{Jacobsen86}. For systems like (TMTSF)$_2$ClO$_4$ (respectively
(TMTTF)$_2$PF$_6$), one finds
$\chi_s/\chi_s^0
\simeq 2 $ (respectively, $\chi_s/\chi_s^0\simeq 3$), which leads to
$U/4t_a\simeq 0.3$ (respectively, $U/4t_a\simeq 0.4 $) 
for a quarter-filled band. 
This simple analysis of the spin susceptibility 
 thus indicates the presence of moderate Coulomb
repulsion. They  are however less than  the values determined 
from elaborate quantum chemistry calculations made for $U$ at the molecular  or
local level
\cite{Ducasse91,Ducasse96},  
 which lead to $U/4t_a \simeq  4-8$. The difference  may be attributed
to the fact that at variance with quantum chemistry  method, the  
Fermi liquid theory  
is a low-energy  theory and the interaction parameter $F^a$
  reflects a screened rather than a local quantity. 

 The Fermi liquid as well as
the Hartree-Fock theory, however,  can hardly account for the observed
temperature dependence of the spin susceptibility obtained at  constant
volume $-$ namely, corrected for thermal dilatation of the sample $-$ 
which shows a monotonic but sizable  growth as  the temperature increases
(Figure \ref{Susceptibility}) \cite{Miljak83,Miljak85,Wzietek93}. This  variation
turns out to be much faster than the one expected  from the H-F theory, 
which only predicts a
$F^a(T)$ with  a  very slow  temperature dependence that is spread out
over $E_F$ on a temperature scale   
 \cite{Wzietek93,Mila95}.  
\begin{figure}[ htb] 
\epsfxsize0.95\hsize
\epsffile{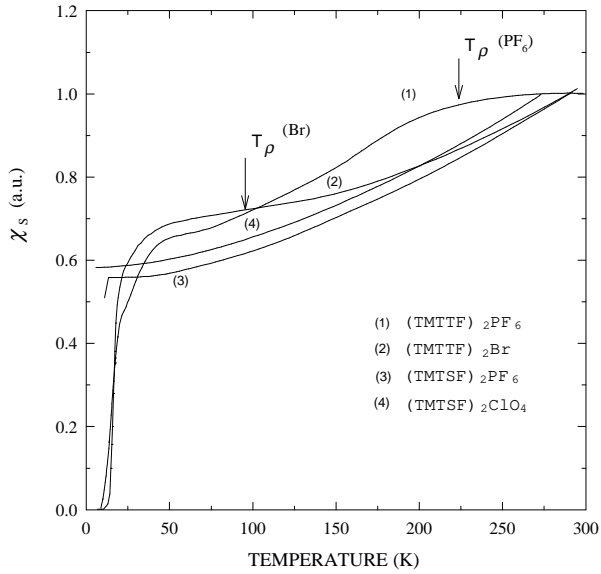}
\caption{Temperature dependence of the magnetic spin susceptibility of members of  (TM)$_2$X
series. The arrows indicate the temperature scale  $T_\rho$ for the onset of the insulating
behavior in the resistivity for (TMTTF)$_2$PF$_6$ and (TMTTF)$_2$Br (see Fig.
\ref{Resistivity}).}    
\label{Susceptibility}   
\end{figure}

\paragraph{Specific heat. $-$}
In
the Fermi liquid theory, the expression for the electronic contribution
to  the specific heat   is linear in temperature $C_e(T) =\gamma T$,
where the Sommerfeld constant is  
\begin{equation}
\gamma= {2\over 3}\pi^2 N(E_F^*). 
\end{equation}
The strength of interaction can also be provided through the Wilson
ratio \cite{Wilson75} 
\begin{equation}
R_W= {\pi^2\over 3\mu_B^2 }{\chi_s\over \gamma}= {1 \over 1 + F^a}.
\end{equation}
 Deviations of $R_W$  from unity thus give information about
the amplitude of interactions.  

The first measurements of  the
temperature dependent specific heat  in the normal state of the Bechgaard
salts   were made by Garoche {\it et al.}
 on  (TMTSF)$_2$ClO$_4$ \cite{Garoche82}.  
 After  subtraction of the phonon contribution, the electronic  part
of the specific heat, albeit restricted to the very low
temperature domain  shows the  metallic   linear behavior $C_e = \gamma T
$ (Figure \ref{Specificheat}).   The  evaluation of the Sommerfeld
constant
$\gamma \simeq 10 $ mJ $\cdot$ mole$^{-1} \cdot $ K$^{-2}$ allows a
determination of the density of states
$N(E^*_F)\simeq 1 $ states/eV/spin/molecule, in fair agreement  with the
one obtained from  the low temperature value of the spin susceptibility
according to the analysis of Miljak {\it et al.} 
\cite{Miljak83}, thereby
lending support to a weak coupling   Fermi liquid picture in this
temperature domain. The analysis of Miljak
\etal  of the spin susceptibility and the electronic
specific  heat   of  (TMTSF)$_2$ClO$_4$  then suggests
that $F^a$ is not large in this system, at least in the low temperature
range. 
\begin{figure}[ htb] 
\epsfxsize0.95\hsize
\epsffile{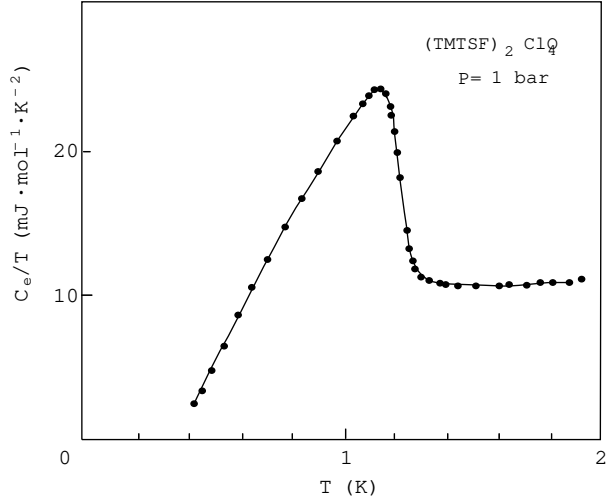}
\caption{Specific heat as a function of temperature in (TMTSF)$_2$ClO$_4$  near the superconducting
transition, after ref.
\cite{Garoche82}. }    
\label{Specificheat}   
\end{figure}

On the other hand, the electronic specific heat reveals an important
field dependence \cite{Brusetti83} (Figure \ref{gamma(H)}). The linear
specific heat coefficient
$\gamma$ increases from 10 mJ$\cdot$mol$^{-1}$ K$^{-2}$ at low field
(although larger than the critical field $H_{c2} = 1
$kOe along $c^\star$), and passes through a maximum of 25 mJ$\cdot$mol$^{-1}$K$^{-2}$ at
$H_m$ = 20 kOe for
$T\approx 1 $ K. The locus of the points corresponding to the maximum of
$\gamma$ in the $ H-T$ plane is located in the paramagnetic domain of
(TMTSF)$_2$ClO$_4$, i.e. not to be confused with the onset of a
field-induced spin-density-wave state detected by the same specific heat
technique \cite{Pesty85}. Hence, the observed decrease of the Wilson ratio
$R_W$  under magnetic field is a peculiar property of this low
temperature electron gas which hardly fits with   the Fermi
liquid model. Furthermore this experimental finding is reminiscent of the
expected behavior in a slightly doped Mott insulator as the
metal-insulator transition is approached (i.e. band filling approaching
half-filling).  Anomalous magnetoresistance is also seen for similar
and larger  field in a broad range of temperature in the normal phase \cite{Behnia95}.
\begin{figure}[ htb] 
\epsfxsize0.95\hsize
\epsffile{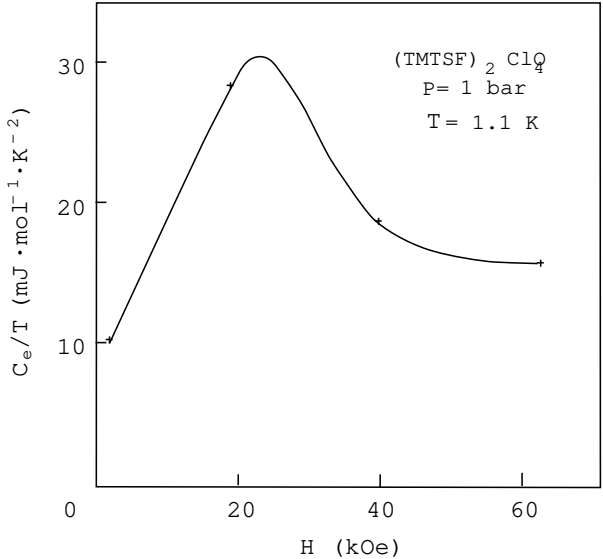}
\caption{Magnetic field dependence of the Sommerfeld constant of specific heat in (TMTSF)$_2$ClO$_4$
at low temperature.  }    
\label{gamma(H)}   
\end{figure}

\subsection{Fermi liquid close to equilibrium: dynamics}
 Key  features of both the non-interacting and the interacting Fermi
liquid theory can be illustrated  through the retarded one-particle
Green function, which can be defined from the following Fourier
transform  
$$
 G_p({\bf k},\omega)= -i\int  e^{i\omega t}\  \langle {\rm GS} \mid
T_t\  a_{p,{\bf k},\sigma}(0)\ a^\dagger_{p,{\bf k},\sigma}(t)\mid {\rm
GS}\rangle  \ dt
$$
taken in the ground state. In a non-interacting Fermi gas, it takes the form
\begin{equation}
 G_p({\bf k},\omega) = {1 \over
\omega-\epsilon_p({\bf k}) -i{\rm sgn }(\epsilon_p)0^+},
\end{equation}
where we have redefined $\epsilon_p({\bf k})-\mu \to \epsilon_p({\bf
k})$. The singularities in the  imaginary part
${\rm Im}G_p({\bf k},\omega)=
\pi\delta(\omega-\epsilon_p({\bf k}))$ indicates that the particle and
the hole states are the true eigenstates of the system. As the
interaction is  turned on there is a one-to-one
correspondence between these non-interacting states and the quasi-particle
states of the  interacting Fermi liquid. This is possible close to the
Fermi surface where the decay rate of quasi-particles is sufficiently
slow compared to the branching time of interaction.
Consequent to this correspondence, the quasi-particles states  are
labeled by the same quantum numbers $({\bf k},\sigma)$ and the coherent part of the 
one-particle  Green function for a Fermi liquid takes the form    
\begin{equation}
  G_p({\bf k},\omega) = {z \over \omega-\epsilon^*_p({\bf
k})  -i{\rm sgn }(\epsilon^*_p)\tau_{\bf k}^{-1}} ,
\label{GreenFL}
\end{equation}
where $z$ is the quasi-particle weight which actually
corresponds to the amplitude of the step of the quasi-particle 
distribution function $n[\epsilon^*_p({\bf k})]$ in a Fermi liquid
($z=1$ in a non-interacting Fermi gas).  Inelastic collisions  introduce
incoherent scattering and an imaginary part in the quasi-particle energy,
which is translated into a decay rate of quasi-particles \cite{Gorkov95,Gorkov96,Yakovenko98},
\begin{eqnarray}
\tau_{{\bf k}}^{-1} \sim  g_3^2 \ {\rm max}\bigl[\omega^2, T^2,
\bigl(\epsilon^*_p({\bf k})\bigr)^2
\bigr]
\label{lifetime}
\end{eqnarray} 
where $g_3$ is
an electron-electron  coupling constant that does not conserve momentum
(Umklapp scattering). The lifetime of quasi-particles 
becomes infinite at  $ {\bf k}_F $  $(\tau^{-1}_{{\bf k}_F }= 0
$ ) due to the  exclusion principle which severely
restricts the possibility of scattering events as we approach the Fermi
surface. Thus at ${\bf k}_F$,  the delta function singularity in the imaginary part
${\rm Im}G_p({\bf k},\omega)=
\pi z \delta(\omega-\epsilon_p({\bf k}))$ indicates that  at the Fermi level quasi-particles become
eigenstates.  

Other  features of a Fermi liquid emerge  when  a perturbation that is slowly
varying  in space and time  is coupled either to  spin or
charge  degrees of freedom.   The mean-field that wraps up each
quasi-particle in a Fermi liquid acts as a coherent
restoring force that leads to a collective response of the system as
a whole.   One  then finds two types of collective or sound modes,
 namely
the charge (or the  plasmons for a metal)  and the spin (paramagnons) 
modes \cite{Pines66}. 

Consider 
for example the paramagnons of an anisotropic 2D Fermi liquid as
described by  the  one-electron spectrum (\ref{flspectre}) with
$t^*_{\perp c}=0 $; 
the imaginary part of the retarded spin response function at low
frequency is found to be  
\begin{equation}
{\rm Im}\chi({\bf q},\omega)= \ \chi_s {\omega_\sigma({\bf
q})\omega\over
\omega^2_\sigma({\bf
q}) + \omega^2}.
\label{spinspectral}
\end{equation}
The absence of pole on the real axis in this expression indicates  
$-$ as it is the case for isotropic Fermi
liquid \cite{Pines66} $-$  strong damping of paramagnons. Nonetheless, the spectral
weight shows a peak at
\begin{equation}
\omega_\sigma({\bf q})= 2(1+F^a)\sqrt{(v_\perp^*q_\perp)^2
-(v^*_Fq)^2}, 
\end{equation}
 which stands as  the dispersion relation of paramagnons where
$v_\perp^*= 2t_{\perp b}^*b$. As a result of the peculiar
anisotropic form of  the electronic spectrum (2), the available  phase
space for 2D paramagnons is reduced so spin excitations are only defined
for $v_\perp^*q_\perp  > v_F^*q$.

\paragraph{NMR nuclear spin  relaxation. $-$}  
 The
temperature  and the magnetic field variation of the nuclear spin-lattice
relaxation ($T_1^{-1}$) as obtained by Nuclear Magnetic Resonance  
 is known 
 to be a quite useful tool for the characterization of 
spin dynamics in  metals \cite{Moriya63,Soda77}. Given the local coupling
between a NMR-active nuclear spin and  the electronic spin density
through the hyperfine coupling, the study of the relaxation process  of
the nuclear spin  gives useful information about the  statics, the
dynamics and the effective dimensionality $D$ of  electronic spin
correlations of different wave vectors $q$ \cite{Bourbon89,Bourbon93}.
This opportunity given by the $T_1^{-1}$ measurements has been recognized
from the start in the study  of the ordered and normal states in  the Bechgaard salts
and their sulfur analogs. The analysis of the nuclear relaxation  essentially focus on the  
following basic expression  due to Moriya \cite{Moriya63} : 
\begin{equation}
T_1^{-1} = 2\gamma_N^2 \mid A \mid^2 T \int d^Dq \ {{\rm Im} \chi({\bf q},\omega)  \over
\omega},
\label{Moriya}
\end{equation}
where $\gamma_N$ is gyromagnetic ratio of the nucleus, $A$ is proportional to the hyperfine matrix
element, and 
${\rm Im}
\chi({\bf q},\omega)
$ is the imaginary part of the dynamic spin susceptibility at wave vector ${\bf q}$ and
Larmor frequency  $\omega$.  
As we have seen, spin fluctuations of an interacting Fermi liquid consist
exclusively of long wavelength damped 
paramagnons which according to (\ref{spinspectral}), are considered
as non-diffusive and for which  
one finds:  
\begin{equation}
T_1^{-1}[{\bf q}\sim 0]=  \ C T \chi_s^{2},
\label{relaxuniform}
\end{equation}
Thus for a 2D Fermi liquid, the enhancement of the nuclear
relaxation with respect to the  non-interacting  `Korringa limit' 
$T_1^{-1} \propto T [N(E_F)]^2 $,  grows as the square 
of the enhancement of the susceptibility \footnote{It should be noted here that the expression for
the relaxation rate  slightly differs  when paramagnons induce  weak ferromagnetism. In  this case,
$(T_1T)^{-1} \sim [\chi_s]^{\dm (5-D)}$ varies as a power of the 
spin susceptibility  whose index  depends on the spatial dimensionality
D \cite{Bourbon89}.}.
 
The experimental situation in  
\sele \ and \sulf\  salts  does reveal an enhancement of
the relaxation rate with respect
 to the  Korringa law which is  compatible with the above expression
at  high enough  temperature (Figure \ref{T1Tchis2}). 
   Detailed investigations of  $\chi_s$ and
$T_1^{-1}$  data for several members of both (TMTTF)$_2$X and
(TMTSF)$_2$X series show that the relation
$T_1^{-1} \propto T[\chi_s(T)]^2$ is apparently well satisfied over 
a large  temperature domain of
the normal phase.  Although this could support the Fermi
liquid theory in this sector of the normal phase, it has been
shown, however, that 1D $-$ undamped $-$ paramagnons of a Luttinger
liquid leads to  a similar expression for $T_1^{-1}$ near $q=0$ \cite{Bourbon89}. 
Moreover, following the example of the temperature dependence of
susceptibility, the FL prediction fails  to account for the
temperature dependence of the enhancement, which according to
Figure \ref{T1Tchis2}, shows a too large  variation
\cite{Wzietek93,Bourbon89}. 
\begin{figure}[ htb] 
\epsfxsize0.95\hsize
\epsffile{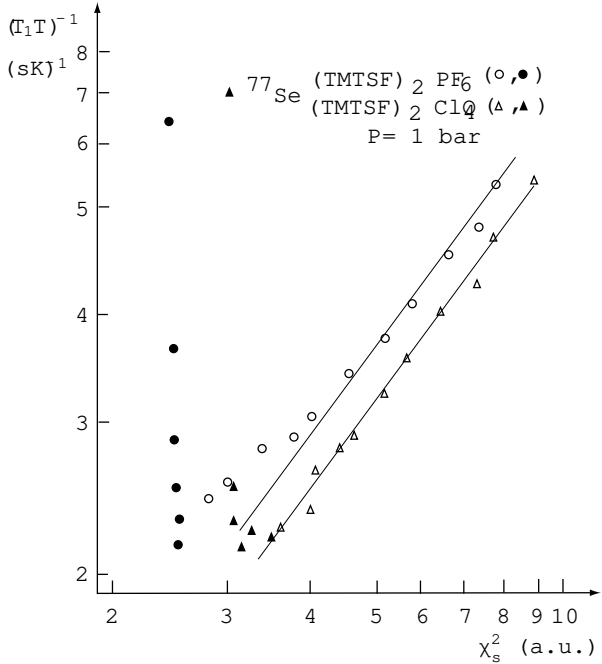}
\caption{$^{77}$Se enhancement $(T_1T)^{-1}$  as a function of the square of the measured
susceptibility for (TMTSF)$_2$PF$_6$ and (TMTSF)$_2$ClO$_4$, after ref. \cite{Bourbon89}.}    
\label{T1Tchis2}   
\end{figure}

The Fermi liquid theory is  more seriously flawed in  the low
temperature part of the normal phase where  pronounced deviations to
(\ref{relaxuniform}) have been  reported  for all compounds studied  in
\sulf\  and \sele\  series. In this temperature range, the variation of
$(T_1T)^{-1}$  shows  an important  increase  instead of the expected decrease and
its occurs  in a temperature region where the  magnetic susceptibility does not show any
appreciable variation.  This enhancement was first reported in the case of (TMTSF)$_2$ClO$_4$
(Figure
\ref{T1SeClO4})
\cite{Bourbon84,Takahashi84,Takigawa86}; it  was afterwards  invariably found in
all members of the (TMTSF)$_2$X and (TMTTF)$_2$X series with more or
less the same profile in temperature depending on the nature of the
ground state. In a compound like (TMTSF)$_2$PF$_6$ at ambient pressure,
the anomalous contribution to the enhancement is found to  grow by a
factor around five between 200\ K and 50\ K that is, well outside the
critical domain associated to the spin-density-wave transition at $T_N
\simeq 12$ K \cite{Bourbon89,Wzietek93};  it is even stronger for members
of the  (TMTTF)$_2$X series where a singular enhancement of the form
$(T_1T)^{-1}\sim T^{-1}$ (or $T_1^{-1}\sim $ constant) is invariably
found (see Figure \ref{T1chisT2})
\cite{Wzietek93}. Emerging very deeply in the normal phase,  this
anomalous behavior of nuclear relaxation  has played an important part
of the debate surrounding  the limitations of the  FL theory  to describe
the normal phase of   these quasi-one-dimensional conductors.  Following
the first observations made on (TMTSF)$_2$ClO$_4$, it was propounded 
that this additional  contribution to the relaxation rate originates in
antiferromagnetic spin fluctuations which are essentially
one-dimensional in character (Figure \ref{T1SeClO4}). 
\begin{figure}[ htb] 
\epsfxsize0.95\hsize
\epsffile{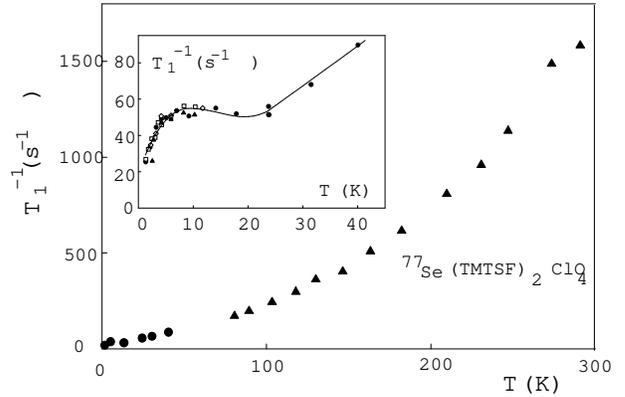}
\caption{$^{77}$Se $T_1^{-1}$ vs $T$ data of (TMTSF)$_2$ClO$_4$, after 
\cite{Wzietek93,Bourbon84}. }    
\label{T1SeClO4}   
\end{figure}    

As the temperature is increased in the normal state the electron lifetime decreases according to
(\ref{lifetime}), it follows that  the dynamics of paramagnons gradually moves away from the
collisionless to the diffusive  limit   when $\omega_e\tau \ll 1$, where
$\omega_e$ is the electronic Larmor frequency.  Probing diffusive spin
dynamics is possible  through the field dependence of the nuclear
spin-lattice relaxation rate; 
being sensitive to dimensionality of the spin system it can
 give
$-$ whenever it is present $-$  quite useful information about the
effective dimensionality of  spin dynamics in organic conductors
\cite{Soda77,Devreux78,Bourbon88b,Azevedo82}. Focusing on the field dependence of the paramagnon
contribution as a function of the spatial dimensionality $D$, one finds 
$$
T_1^{-1} \sim  (\omega_e\tau)^{-\dm} \ \ \  (D=1),
$$
$$
T_1^{-1} \sim  \ln{1\over \omega_e\tau} \ \ \ \ \  (D=2),
$$
\begin{equation}
 T_1^{-1} \sim  {\rm constant}  \ \ (D=3).
\label{diffusive}
\end{equation}
The search for a field  dependence of the nuclear relaxation rate in the normal phase of
(TMTSF)$_2$ClO$_4$  has been set out by Caretta {\it et al.}
\cite{Caretta95}. Their results obtained at 200~K (Figure
\ref{T1ClO4H})  up to 15 Tesla show a  square-root field dependence
which is indicative  of one-dimensional diffusive spin dynamics. The
square root dependence is cut off  at low enough field due to
processes that do not conserve spin along the stacks \cite{Soda77}. 
The  field dependence is found to become essentially undetectable  around
150~K and below, which may indicate  either a change of dimensionality in the spin
dynamics or that the collisionless regime is reached. 
\begin{figure}[ htb] 
\epsfxsize0.95\hsize
\epsffile{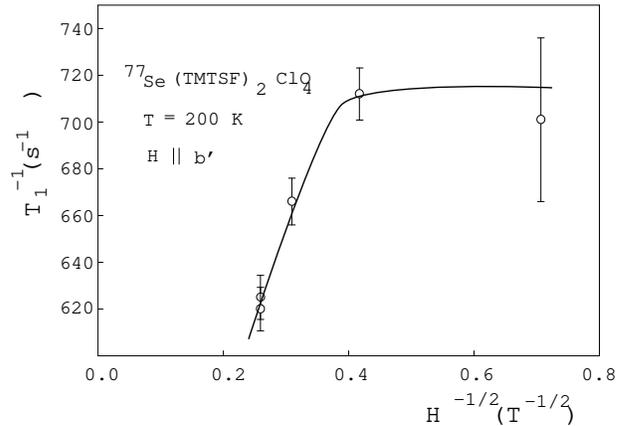}
\caption{Field dependence of the $^{77}$Se  nuclear spin-lattice relaxation rate in
(TMTSF)$_2$ClO$_4$, after \cite{Caretta95}.  }    
\label{T1ClO4H}   
\end{figure}

Earlier investigations of Azevedo {\it et al. }\cite{Azevedo82}, obtained on $^1$H
of (TMTSF)$_2$PF$_6$ in the high-pressure metallic phase at  very low temperature revealed the
existence of a field dependence of 
 $T_1^{-1}$ at 4~K  up to  a field of 10~kOe, which has been held to
 fit in with the logarithmic
profile expected for  diffusive two-dimensional spin dynamics (Eq. \ref{diffusive}).

\paragraph{Transport: (TMTSF)$_2$X. $-$} 

In a metallic system where  phonons and impurities  play little role
in the scattering rate
of  carriers,  the temperature dependence of  resistivity   is 
controlled by electron-electron scattering rate. In a Fermi liquid, 
it is then governed by the  decay
rate of quasi-particles. Dropping the logarithmic factors due to either the special geometry of the
Fermi surface or the proximity of the SDW transition, one gets a   quadratic
temperature dependence of the form
\cite{Gorkov96,Gorkov98,Yakovenko98}:
\begin{equation}
\rho(T)
 \propto \tau_{{\bf k}_F}^{-1}\sim  T^2. 
\end{equation}
 Within the Fermi liquid picture this temperature dependence for the
resistivity should hold for at least two spatial directions at $T\ll
T_{x^1}$. 

 Before
making any comparison  with actual data in systems like \sele,  one
should be aware of the remarkably strong pressure (or volume) dependence
of the transport properties. The longitudinal conduction  of
(TMTSF)$_{2}$X increases at a rate of 
$\approx 25\%$ kbar$^{-1}$.  This pressure coefficient is significantly
larger than the figure expected for a scattering mechanism governed by
acoustic phonons in the Boltzmann formalism since the pressure
dependence of the bare bandwidth is of order $2\%$ kbar$^{-1}$ in 1-D
organic conductors as measured either from the pressure dependence of the
longitudinal plasma edge \cite{Welber78}, or from the pressure
dependence of the Fermi wave vector in TTF-TCNQ \cite{Megtert81}. 

Owing to  the significant pressure dependence of $\rho_{a}(T)$  observed down
to low temperature, the temperature variation  of the resistivity measured
under constant pressure contains two contributions.  One is coming from
the temperature dependence of the scattering processes and another one
is related to the volume dependence of these processes through the
thermal contraction of the lattice. In order to obtain experimental data
which can be confronted to the theory the observed temperature dependence
under constant pressure must be transformed into a constant volume
dependence. As some arbitrariness remains in the conversion procedure, the
converted data must be considered at best as an improvement before
comparison with the theory is made. The unit cell of the crystal at 50~K
has been taken as the reference unit cell (such an hypothesis being
justified by the lack of thermal expansion at this temperature). The
procedure goes as follows. At a temperature $T$ above 50~K, a pressure
$P$ is needed in order to recover the reference volume. An important
simplification of the procedure has been made assuming that the cell length
along $a$ is the most relevant parameter (instead of the volume) to be
taken into account. Hence, thermal expansion, compressibility data and
isobaric temperature dependence of the resistivity enable a point by point derivation of the
temperature dependence at constant volume which is displayed in Fig. \ref{ResistV}.
\begin{figure}[ htb] 
\epsfxsize0.95\hsize
\epsffile{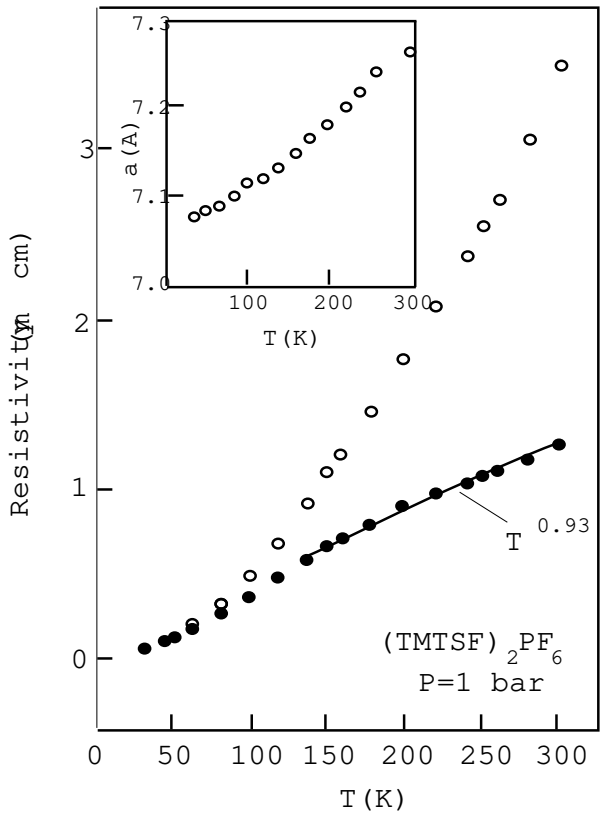}
\caption{Longitudinal resistivity of (TMTSF)$_2$PF$_6$ as a function of temperature at constant
pressure (open circles, Ref. \cite{Auban99}) and converted to constant volume (full circles). In the
inset, the variation of the stack lattice parameter with temperature  after \cite{Gallois87}. }    
\label{ResistV}   
\end{figure}

This conversion procedure must be attempted for the
longitudinal as well as for the  transverse  transport. The results for
$\rho_a(T)$ in  (TMTSF)$_2$PF$_6$ at ambient pressure is given in
Fig.~\ref{ResistV} where a crossover from a superlinear to a linear (or
sublinear) power law temperature dependence is observed in the vicinity
of $80$ K. A detailed  analysis based on the Fermi liquid theory 
shows that a
$T^2$ law is apparently well satisfied below 50 K down
to the vicinity of the spin-density-wave transition where the resistivity
is dominated by critical scattering effects \cite{Gorkov98}.

The resistivity measurements along the transverse $ c $ direction  are
also interesting. Actually one can infer that $\sigma_c$  is directly
related to the physics of the
$a-b$ planes and therefore could probe whether transport proceeds via
collective modes or independent quasi-particles. Jacobsen {\it et al.} \cite{Jacobsen81b}, were the
first to report a non-monotonic temperature dependence of  $\rho_c$ in
(TMTSF)$_{2}$PF$_6$ passing through a well characterized maximum at
$T_{m} = 80$ K (under ambient pressure) `at variance' with the
$T$-dependence in the a direction \cite{Jacobsen81b}. A recent
pressure study of this effect has shown that  $T_{m}$ evolves under
pressure and reaches about $300 $ K at
$10$ kbar \cite{Moser98}. The constant volume data for $\rho_c(T)$ in the `metallic'
regime below $T_m$ reveals that $\rho_c(T)\propto T^{1.5}$, which
differs from a $T^2$ law and may indicate that the transport along
the less conducting axis is incoherent and diffusive since this
temperature region is still characterized by $T \gg t_{\perp c}^*$.  
As for DC resistivity measurements along $b$ they are more difficult to
be neatly realized owing to  non-uniform   current distributions between
contacts which introduce contributions coming from other directions.
Nevertheless, if we apply a tunneling argument between the  $a-b$ planes
instead of chains that is, if quasiparticle states with mean free paths
of order 
$\tau_{a}t_{a}/\hbar$  and  $\tau_{b}t_{\perp b}/\hbar$ can be defined along
$a $ and $b $ directions respectively, the transverse conductivity reads
$\sigma_{c}\approx (\sigma_{a}\sigma_{b})^{1/2}$ \cite{Moser98}. Hence, provided the
quasi-particle life times along $a$ and
$b$ exhibit the same temperature dependencies the anisotropy ratio
$\rho_c/\rho_a$  should be $T$-independent. This is definitely not
observed below $T_{m}$ as experimental laws such as $\rho_{a}(T)
\approx T^{2}$ and $\rho_{c}(T) \approx T^{1.5}$  are more appropriate
leading to $\rho_{b}(T) \approx T^{}$  within the quasi-particle 
hypothesis for the
$a-b$ planes. A linear temperature dependence is indeed very close to the
early experimental findings for
$\rho_{b}(T)$  below  $T_{m}$ under ambient pressure suggesting that
the quasi-particle states if they can be defined at all are not
Fermi-liquid-like but possibly marginal. 
\begin{figure}[ htb] 
\epsfxsize0.95\hsize
\epsffile{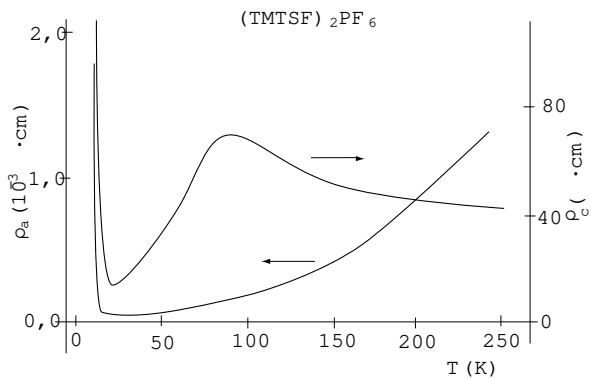}
\caption{Longitudinal and transverse $c$ resistivities as a function of temperature in
(TMTSF)$_2$PF$_6$, after \cite{Moser98}.  }    
\label{resistance-ac}   
\end{figure}

Very recently,   Fertey {\it et al.} \cite{Fertey98}, showed that a direct 
measurement  of
$\rho_b(T) $ on a system like  (TMTSF)$_2$PF$_6$ is possible using
the microwave technique and the results show the existence of a maximum
of  $\rho_b(T) $ as a function of temperature  around 40~K $-$ which
is somewhat lower  than the value of $T_m$ shown by  
$\rho_c(T)$. 

The striking different temperature dependencies for the {\it
in} and {\it out} of plane resistances suggest an interpretation in terms
of a non-Fermi liquid approach  for $T > T_{m}$, although
$\rho_{a}(T)$ is not far from a conventional behavior.  The Figure \ref{resistance-ac} emphasizes
the remarkable feature of
(TMTSF)$_{2}$PF$_6$, namely, opposite temperature dependencies for
interplane and chain resistivities above
$T_{m}$.
A temperature independent
anisotropy ratio is indeed observed below $10$~K in (TMTSF)$_2$PF$_6$
under 9 kbar,  suggesting a recovery of the usual Fermi
liquid behavior ( Figure \ref{anisotropy}).
\begin{figure}[ htb] 
\epsfxsize0.95\hsize
\epsffile{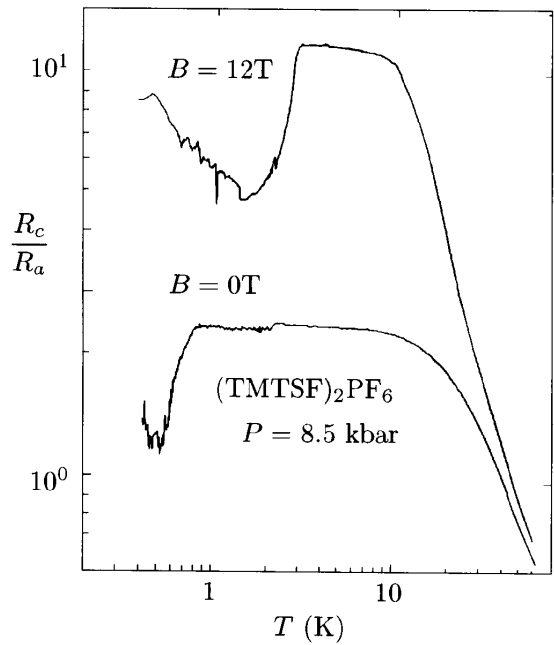}
\caption{Anisotropy ratio for the resistance along the $a$ and $c$ directions in
(TMTSF)$_2$PF$_6$,  after\cite{Jerome99}.  }    
\label{anisotropy}   
\end{figure}

\paragraph{Transport:~\sulf.$-$}

Deviations to the Fermi liquid behavior are  particularly  revealing for (TMTTF)$_2$X
compounds for which  there is a lost of the metallic character  at a characteristic temperature 
$T_\rho$ that  can be in magnitude  far  
  above  the critical temperature domain under  low pressure conditions (Figure \ref{Resistivity}).
The  scale
$T_\rho$  signals a  thermal activation of  carriers  corresponding to a gap
$\Delta_\rho \sim 2T_\rho$ in  the electrical resistivity of the normal phase
 .
 The recent data give $2\Delta_{\rho}\approx 900 K$ under
ambient pressure \cite{Moser98,Brown97}, which only slightly differs from
previous estimations \cite{Laversanne84}. 

The comparison with the temperature profile of the spin susceptibility 
is rather striking since the spin susceptibility remains unaffected
despite the thermal activation of carriers  below 
$T_\rho$ (see Figures \ref{Resistivity} and \ref{Susceptibility}). This apparent decoupling
between spin and charge degrees of freedom  is in severe contradiction
with what should be  expected  in a  Fermi liquid.

\paragraph{Optical properties.$-$} 
Optical reflectivity measurements performed by  Jacobsen {\it et
al.} \cite{Jacobsen81,Jacobsen82,Jacobsen83}, were probably the first to provide some direct 
information about the anisotropy  of the electronic structure in
the Bechgaard salts (Figure \ref{RefelctanceJacob}). The observation of
the growth of an infrared reflectance edge in the transverse $b'$
direction of (TMTSF)$_2$PF$_6$ below 100~K revealed the presence of a
sizable overlap integrals in that direction. Thus the sharpness of a
Drude like  plasma edge in the $b'$ direction lends support for the
gradual emergence of a coherent transverse one-particle motion and a 2D Fermi
liquid component in the $a-b$ plane in the temperature region below 100~K.
Similar results were subsequently found for other members of the
(TMTSF)$_2$X series and  the analysis of the  reflectivity data using
a tight-binding spectrum (\ref{spectrum}) yields $t_{\perp b }\simeq
18-22 $~meV and $t_{\perp b}/t_a\simeq .1$ for the highest transverse
hopping integral and the anisotropy ratio \cite{Kwak82,Yamaji90}. These
values were found to be in a reasonable agreement with the band
structure  calculations. This would also indicate a small
renormalization of $t_{\perp b}$, namely $t_{\perp b}^* \to t_{\perp
b}$ in the energy range of the plasma edge. However it is worth
noting that the transverse plasma edge could be insensitive to the
quasi-particle renormalization factor $z$ in which case, it would only
reflects the bare unrenormalized band parameters. 
\begin{figure}[ htb] 
\epsfxsize0.95\hsize
\epsffile{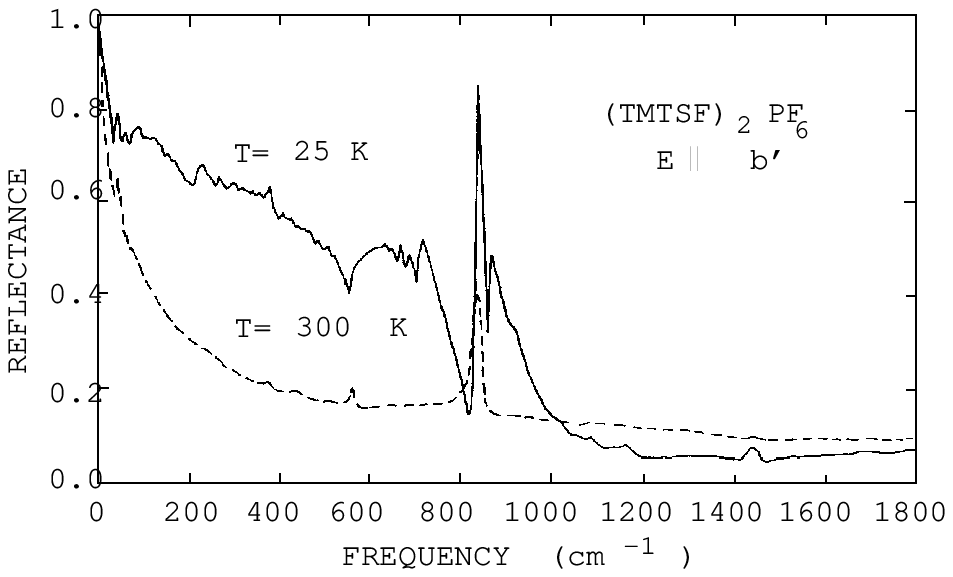}
\caption{Polarized refelectance for  {\bf E}$\parallel b'$ of (TMTSF)$_2$PF$_6$ at $T=25$K
and 300~K, after \cite{Jacobsen83}.  }    
\label{RefelctanceJacob}   
\end{figure}

In contrast, no plasma edge is found along the $b$ and $c$ directions in (TMTTF)$_2$X 
due to the presence of an insulating state. In this case the  charge motion perpendicular to the
chains is   apparently absent and the  presence of a charge gap makes the single particle hopping
$t^*_\perp$
  irrelevant at low temperature \cite{Bourbon95,Suzumura98,Vescoli98}.

The frequency dependence of  optical conductivity  obtained  by  Kramers-Kronig transforms of
the reflectance  data, however, can hardly   bow to an analysis based on
the Drude model \cite{Timusk96,Dressel96}. Despite the
 large value of  DC conductivity in (TMTSF)$_2$X materials at low temperature, the  
amplitude of  $\sigma(\omega)$ in the far infrared is much smaller than expected within a simple
Drude picture (Figure \ref{SigmaTimusk}). The absence of a classical
Drude peak in the Bechgaard salts  was actually reminiscent to what was
previously observed in the case of incommensurate CDW system TTF-TCNQ. 
It was  proposed that the  narrow zero-frequency peak in conductivity is
the result of a collective mode effect that bears some similarity with
the  sliding a charge-density-wave $-$ especially if one considers the new phonon features that
emerge at low temperature  
\cite{Timusk96}. This is all the more surprising if one considers that (TMTSF)$_2$X compounds are
very good metals in their  the normal state and that the ground state of these materials is either
 antiferromagnetic or  superconducting.   
 Further, the zero frequency mode is also accompanied by a gap ranging from  160 cm$^{-1}$ to 200
cm$^{-1}$ depending on the compounds. This anomalous feature has been reported by several
groups \cite{Jacobsen81,Timusk96,Dressel96,Vescoli98}. It was recently proposed
to be associated to the remnant of a  Luttinger liquid with a correlation
gap  which is present in the less metallic  sulfur series (see the 
section on the Luttinger liquid)\cite{Vescoli98,Giamarchi97}.

 A
different, albeit one-dimensional,  interpretation due to Favand and 
Mila
\cite{Mila96}, is based on the fact that the energy associated to the
dimerization gap falls in the same  range of the infrared spectrum. 
Combined to strong local repulsion among carriers which spreads the
single
 occupancy of the electron states to the band edge, this would lead to a
relatively pronounced absorption  due to interband transitions
\cite{Pedron94}. In this scenario, the  infrared absorption  
conductivity shifts to lower frequency when the dimerization
gap  decreases  
 as one moves to the right in the phase diagram (Figure
\ref{Diag-phase}). 
\begin{figure}[ htb] 
\epsfxsize0.95\hsize
\epsffile{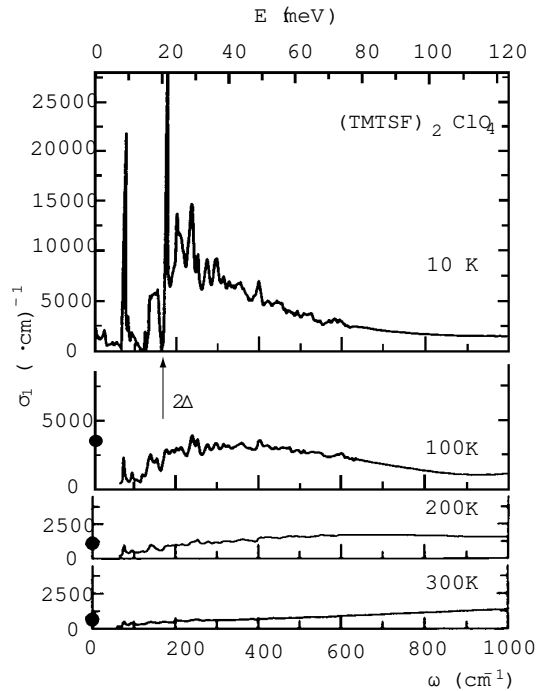} 
\caption{Real part of
$\sigma(\omega)$ of (TMTSF)$_2$ClO$_4$  along the chain axis at  different temperatures in the
normal phase. The full circles on the left are the values of DC conductivity. At 10 K, there is a
gap with the value 2$\Delta\simeq $ 170 cm$^{-1}$. After \cite{Timusk96}.}    
\label{SigmaTimusk}   
\end{figure}

	A very strong support in favor of an insulating state in \sulf \ can also be 
given by the optical data \cite{Jacobsen83}. The reflectivity spectra
reveals the existence of a Drude edge for the light polarized along the
$a$-direction leading to
$\omega_p = 8860\  {\rm cm}^{-1}$ \cite{Jacobsen83}. A  Drude model fit
based  on the band structure (\ref{spectrum}) or
(\ref{flspectre}) relates the plasma frequency to the band width
and leads to either 
$W_a = 280$~meV  or  $800$~meV depending if a  1/2-filled or  1/4-filled
band interpretation  is adopted. The former assumption is in fairly good
agreement with the ab-initio calculation for (TMTTF)$_{2}$PF$_6$ (see
Table I).  The optical investigation has been extended recently towards
the FIR range showing a leveling-off of the reflectivity at
$80\%$ below a plasma frequency  $\omega_p = 6000\ 
{\rm cm}^{-1}$ \cite{Vescoli98}. No plasma edge is observed at any
temperature when the light is polarized along the
$b$-direction \cite{Jacobsen83,Vescoli98}. In terms of
$\sigma(\omega)$, an insulating behavior is observed with an optical gap
of $\approx 800\  {\rm cm}^{-1}$ in
(TMTTF)$_{2}$PF$_6$.  The conductivity above the optical gap of 800 
cm$^{-1}$ contains almost all the spectral weight corresponding to 
 $\omega_p =
6000\  {\rm cm}^{-1}$ namely, the sum rule condition 
$$
\int_{800}^{\infty} \sigma(\omega) d\omega =
\omega_{p}^{2}/8,
$$ 
is apparently well satisfied \cite{Vescoli98}.


\begin{figure}[ htb] 
\epsfxsize0.95\hsize
\epsffile{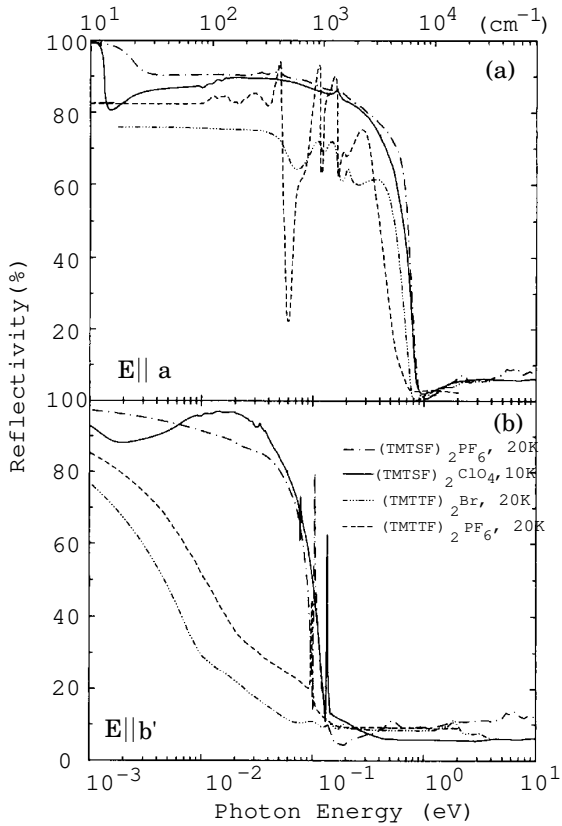}
\caption{Optical reflectivity measurements for electric field parallel (a) and perpendicular to
chains (b). After ref. \cite{Vescoli98}.}    
\label{RefelctanceVescoli}   
\end{figure}

\paragraph{Photoemission.$-$}  
In  experiments  such as the angular resolved photoemission spectroscopy
(ARPES), one has direct access to the momentum and energy dependent
one-particle spectral density
\begin{equation}
A({\bf k},\omega)= {\rm
Im}G({\bf k},\omega),
\end{equation}
 which is defined as the  imaginary part of the
one-particle Green function $G({\bf
k},\omega)$ \cite{Grioni97}. In a Fermi liquid for example, $A({\bf
k},\omega)$ is 
the  probability that at a given energy $\omega$, it exists a
quasi-particle state at
$\epsilon({\bf k})$. It  then expected to show a dispersing peak at
$\omega=\epsilon({\bf k})$, whose width equals $
\tau_{\bf k}^{-1}$ and goes in principle to zero at $T=0$K $-$ albeit in
practice, it is  limited to the experimental  resolution in
energy. For a Fermi liquid, $A({\bf k}_F,\omega=\mu)$ should be finite at
the Fermi level. So far, however, photoemission
experiments have failed 
 to detect any sign of quasi-particle states in both
\sele\ and \sulf\  series \cite{Dardel93,Grioni97}.

The first photoemission experiments were made by Dardel
{\it et al.} \cite{Dardel93}, on (TMTSF)$_2$PF$_6$ at 50~K and the
results were quite puzzling since  no quasi-particle weight in the
momentum-integrated intensity
${\rm Im Tr}_{k_a}G({\bf k},\omega)$ was found at the Fermi edge
- this quantity corresponds  to the density of states.  Further the photoemission signal raises as
a power of the energy from the Fermi edge and forms a  broad but non-dispersing
peak centered at a high energy of about  1~eV below
$E_F$. These anomalous features
of the ARPES spectra remain largely unexplained.
\begin{figure}[ htb] 
\epsfxsize0.95\hsize
\epsffile{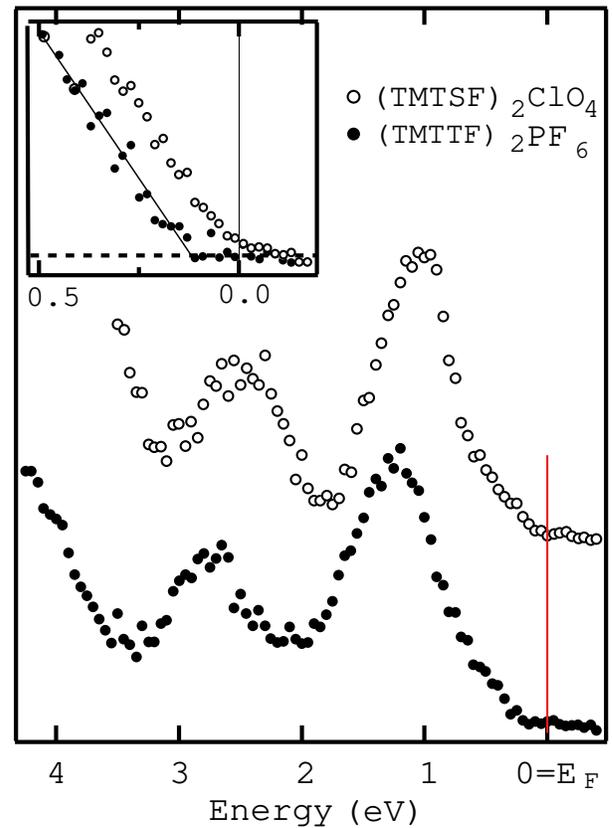}
\caption{ARPES spectra of (TMTSF)$_2$PF$_6$ and (TMTSF)$_2$ClO$_4$ at $\Gamma$ point (cf. Figure
\ref{bande}). The inset identifies an energy shift compatible with a charge gap in
(TMTTF)$_2$PF$_6$. After \cite{Grioni97}.  }    
\label{photoemission}   
\end{figure}

The ARPES spectra  signal for the sulfur compound (TMTTF)$_2$PF$_6$,
 displays   a rigid shift of the leading edge near the Fermi energy to
 about 100~meV. This value is consistent with the
charge gap of 900 K obtained from transport
experiments   DC  or 800 {\rm cm}$^{-1}$ from optical  conductivity 
of (TMTTF)$_2$PF$_6$.  
 Within a one-dimensional  frame of interpretation, this gap has been  ascribed to a
Mott-Hubbard localization gap  \cite{Grioni97}.

\paragraph{The Fermi liquid theory and the SDW instability.$-$ }
Although the present review is not really concerned with  the nature
and the mechanisms of long-range ordering taking place in (TM)$_2$X,
these  played, though indirectly, a relevant role when one tries to assess
the viability of the Fermi liquid theory in spin-density-wave systems like the \sele. Following the
example of  the success of the  BCS mechanism    to explain the onset of conventional
superconductivity as an instability of  a Fermi
liquid, 
there are many features of the spin-density-wave  phase transition either in  
\sele\ at low pressure or in \sulf\  at high pressure that can be
 fittingly described assuming the existence of a well defined  Fermi
liquid component in the normal phase. This can be  substantiated   if one looks at the  mechanism  
 of suppression of antiferromagnetism under
 pressure  in the Bechgaard salts (Figure \ref{Diag-phase})
\cite{Yamaji82,Seidel83}.  It is generally held that
the rapid drop of  the  SDW $T_c$ under pressure   
 (Figure \ref{Diag-phase})  results from   the gradual  frustration  of
a particular symmetry property of the 
electronic spectrum $-$  the so-called nesting.  For an open  Fermi surface as in the Figure
\ref{bande} or the one resulting from the spectrum (\ref{flspectre}), perfect nesting
conditions  prevail so that   electron and hole states on
opposite sides of the  Fermi surface are connected through the relation
\begin{equation}
\epsilon_p({\bf k}) 
= -\epsilon_{-p}({\bf k} + {\bf Q}_0),
\label{symmetry}
\end{equation} 
where ${\bf Q}_0=(2k_F^0,q_\perp^0)$ is the nesting vector that
perfectly maps one part of the Fermi surface on the other.

Accordingly, the electron gas  develops a singular logarithmic
response 
$$
\Pi^0({\bf Q}_0) \sim  \sum_{\bf k} {n[\epsilon_p({\bf
k})]-n[\epsilon_{-p}({\bf k} + {\bf Q}_0)]\over  \epsilon_p({\bf k}) 
 -\epsilon_{-p}({\bf k} + {\bf Q}_0)}
$$
\begin{equation}
\sim N(E_F^*)\ln {E^0_x\over  T},  \hskip 1.3 truecm
\label{nesting2D}
\end{equation}
to the electron-hole pair or density-wave formation at ${\bf Q}_0$,
where $E^0_x $ is a cut-off energy.   
A repulsive electron-electron coupling $\lambda $ leads to an effective attraction between an
electron and a hole  separated by ${\bf Q}_0$. The   coupling  to the singular electron
gas response through a ladder diagrammatic summation  will then
predict an instability of the normal state
$-$ preferentially of the spin-density-wave type
$-$  when
$g\Pi^0
\sim 1$, that is for 
\begin{equation}
T_c \sim E_x^0 e^{-2/\lambda}.
\end{equation} 
In practice, however,  the electron-hole symmetry relation (\ref{symmetry})  is never perfectly 
satisfied. There are  deviations due to small corrections neglected in the  spectrum
(\ref{flspectre}).  Following a lattice compression, the electron spectrum is altered and these
deviations   magnify  under pressure and tend to suppress the logarithmic
singularity (\ref{nesting2D});  $T_c$ thus rapidly decreases  and even vanishes above some critical
pressure (Fig. \ref{Diag-phase}).
    
A prerequisite which is at the  heart of this nesting mechanism  is the existence of a coherent 
Fermi liquid  component in at least two spatial directions for temperatures below $E_x^0\sim T_{x^1}
$. Actually,  the possibility to single out the electron-hole scattering channel in
the ladder summation rests on the existence of quasi-particles and a coherent 2D
Fermi  surface $-$ in the same way as the
electron-electron pairing channel is selected in the BCS theory of  superconductors. Otherwise for
$T > t^*_{\perp b}$, the system is effectively 1D in character and there is no possibility to select
the electron-hole pairing channel. The electron system turns out not to be   a Fermi liquid (see
Section 4).

In the same vein, another example  one can quote 
in support of a Fermi liquid component  in the right-hand side of the phase diagram 
is the unequivocal  success of a description  `\`a la BCS' of the
field-induced-spin-density-wave state  phenomena taking place at very low temperature in these
compounds
\cite{Chaikin96,Gorkov84,Heritier84,Yamaji86,Chen86}. This instability of the normal phase occurs in
the presence of nesting frustration. A magnetic field applied along
the $c$ direction, however,  restricts the electron motion along the $b$ direction
and `one-dimensionalizes'  the kinetics of quasi-particles. Perfect but quantized 
nesting conditions are  found to be restored which lead above some   threshold field to an
instability of the Fermi liquid  toward a  cascade of spin-density-wave states.  

\section{The quasi-one-dimensional approach to the normal phase of the
Bechgaard salts} 
The  point of view according to which the influence of
low-dimensional physics is much more  expanded  in the normal phase of Bechgaard salts corresponds
to  the situation where the scale
$T_{x^1}$ for the  coherence in the transverse one-particle motion  is at much lower temperature
and  can even become irrelevant under particular conditions. This  naturally compels to  call upon
one-dimensional  models of interacting electrons  whose low-energy properties are known to  differ
from those of  a Fermi liquid. 

\subsection{Electron gas model}
In order to  bring out the essential features behind the Luttinger
liquid state we consider the electron gas model  \cite{Dzyaloshinskii72}. This model  is of course 
 a crude simplification of the  interacting electron structure that takes place in actual organic
compounds.  Being defined with a small number of  parameters, the model  is nevertheless 
generic of a different  kind of an electronic state that might be relevant to the description of
organic conductors.

The 
model is based  at the outset from the observation that a 1D non-interacting Fermi gas 
with a two-point Fermi surface  at 
$\pm k_F^0$ gives rise to two  infrared logarithmic singularities in the response of 
the system to correlate pair of particles either in  the
electron-electron (Cooper) or in the $2k_F^0$ electron-hole (Peierls)
scattering channel. These are of the form 
$$
\Pi^0 \sim 2N(E_F) \int_T^{E_F} {d\epsilon\over \epsilon(k)- m\epsilon(k')}
$$
\begin{equation}
  \sim N(E_F) \ln {E_F\over T}.
\label{log}
\end{equation}     
 The singularity, though similar for both types of pairing, results from a distinct symmetry
property of the one-particle spectrum. Take for example the
logarithmic integration  of the Cooper channel where $m=-1$ and $k'=-k$; it
express the formation of pairs of particles of total momentum zero for
which
$\epsilon(k)=\epsilon(-k)$ is satisfied for the particles of  each pair. In three
dimensions, the slightest attraction between electrons in
these time-reversed states inevitably leads to an instability of the
metallic state towards BCS type of superconductivity. In one  dimension,
however, this singularity is not alone since the electron spectrum 
displays another type of symmetry property called `nesting' for which $m=1$ and $k'=k-2k_F^0$, where
this time the energies  of an electron (or a hole)  state at $k$ and a hole (or an
electron ) state  at
$k-2k^0_F$ are  connected  through the relation
$\epsilon(k)=-\epsilon(k-2k^0_F)$ $-$ the 1D analog of eq. (\ref{symmetry}). The summation over  a
macroscopic number of  intermediate states that are connected by nesting   is  responsible for  a
logarithmic integration of the form (\ref{log}). It is the underlying mechanism
of  the  Peierls lattice  instability when  $2k^0_F$ electron-hole pairs are coupled to
low-frequency phonons \cite{Peierls55}. 

 What thus really makes one dimension so peculiar resides in the fact
that the  symmetry  of the spectrum for the Cooper and Peierls instabilities
refer to the same phase space of electronic states \cite{Bychkov66}. The two different
kinds of pairing act as independent and simultaneous processes of the
electron-electron scattering  amplitude  which interfere  with and
distort each other  at all order of perturbation theory.   What comes out  of
this interference is  neither a BCS superconductor nor a Peierls/density-wave
superstructure but a different  instability of the Fermi liquid called a 
Luttinger liquid. 

Following the example of a Fermi liquid, however,  a selected emphasis is
put  on the electronic states  close to the Fermi level where the argument
$E_F/T$ of the logarithm goes to infinity at $T\to 0$. In the framework
of the electron gas model, this amounts  to write the one-particle 
kinetic term of the Hamiltonian in the form
\begin{equation}
H_0 = \sum_{p,k,\sigma} \epsilon_p(k) \
a^\dagger_{p,k,\sigma}a_{p,k,\sigma},
\label{hamiltzero}
\end{equation}
where $\epsilon_p(k)\simeq v_F(pk-k_F^0)$ is  linearized around
the two Fermi points $p k_F^0$ for right $p=+$ and left $p=-$ moving
electrons (a similar continuum approximation neglecting the details at
short distance has been made  in the context of the Fermi
liquid theory [cf. Eq.(\ref{flspectre})]).

 Accordingly,
direct interactions between carriers  in the model can be defined  close
to the Fermi points, a  procedure called the `g-ology' decomposition of
the interaction
\cite{Dzyaloshinskii72,Solyom79}. For a rotationally invariant system,
it allows to single out four different coupling constants:  the
backscattering and the forward scattering terms
$g_1$ and $g_2$, for which two electrons near opposite Fermi points are
coupled through a  momentum transfer near $2k_F^0$ and
zero, respectively; the $g_3$ coupling corresponds to Umklapp
scattering  where two electrons near $+k_F^0$ (resp. $-k_F^0$) are
backscattered  to $-k_F^0$  (resp. $+k_F^0$), a process  that is made
possible at half-filling if the reciprocal lattice vector
$G=4k_F^0=2\pi/a $ enters in the momentum conservation law; finally, one
has the coupling $g_4$ by which two electrons near $k_F^0$ (resp.
$-k_F^0$) experience a small momentum transfer which keeps them on the
same branch \cite{Solyom79} (Figure \ref{gology}). 
\begin{figure}[ htb] 
\epsfxsize0.95\hsize
\epsffile{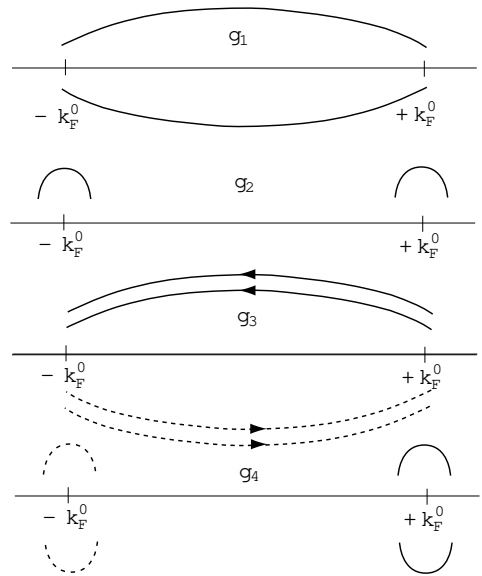}
\caption{g-ology decomposition of the electron-electron interaction close to the Fermi points $\pm
k_F^0$ in the electron gas model. }    
\label{gology}   
\end{figure}    

Focusing for the moment on the most important couplings $g_{1}$, $g_{2}$ and $g_{3}$, the
interacting part 
$H_I$ of the total  Hamiltonian  can be written in the form
$$
H_I = (2g_2-g_1) \sum_{p,q} \rho_{p}(q)\rho_{-p}(-q) -
g_1\sum_{p,q} {\bf S}_{p}(q) \cdot {\bf S}_{-p}(-q) 
$$
\begin{equation}
   +{1\over 2L}\sum_{\lbrace p,k, Q,\sigma \rbrace}g_3\ 
a^{\dagger}_{p,k_1+pQ,\sigma}a^{\dagger}_{p,k_2-pQ+pG,\sigma'}
a_{-p,k_2,\sigma'}a_{-p,k_1,\sigma},  
\label{hamiltonian}
\end{equation}
where 
$$
\rho_p(q)= \dm (L)^{-\dm}\sum_{\{k,\alpha\}}a^\dagger_{p,\alpha}(k+q)a_{p,\alpha}(k),
$$
 and
$$
{\bf S}_p(q) = \dm (L)^{-\dm}\sum_{\{k,\alpha\beta\}}
a^\dagger_{p,\alpha}(k+q)\vec{\sigma}^{\alpha\beta} a_{p,\beta}(k),
$$ 
are respectively 
 the long wavelength  
charge- and spin-density operators   for  right- or  left-moving
carriers.

The connection of the effective intrachain g-ology couplings with the
intramolecular or one-site repulsive interaction  and the screened long-range
Coulomb matrix element have been discussed in detail by Barisic {\it et
al.} \cite{Barisic83,Barisic84,Barisic85}. For $g_1, g_2 $ and $g_3 $, one has the following
result
\begin{eqnarray}
&&g_1
\approx Ua,\cr
&&g_2 \approx Ua + 2Va(\ln{E_F\over \omega_p} +1),
\label{couplings}
\end{eqnarray}
where $\omega_p$ is the plasma frequency and $V$ is the Coulomb matrix element  between neighboring
 molecular sites.  The screening $-$ here non-logarithmic $-$ of the  long range part of the Coulomb
interaction thus appears in the renormalization of the forward scattering amplitude.
As  for the Umklapp term $g_3$, if one refers exclusively to the
amplitude of the dimerization in the evaluation of the half-filled
character  of the band \cite{Barisic81,Seidel83,Mila94}, one finds in the lowest order 
\begin{equation}
g_3 \approx g_1 {\Delta_D\over E_F},
\label{g3}
\end{equation}
where $\Delta_D$ is the dimerization gap at $\pm 2k^0_F$. This expression
does not account, however, for the full influence of the anion lattice
on the amplitude of Umklapp scattering \cite{Emery82,Emery83}. 
Although a $4k_F^0$ anion potential in the presence of  an infinitely rigid
lattice  produces no dimerization, 
the $4k_F^0$ bond-charge modulation induced by the anions subsists,
however, and is sufficient to enhance Umklapp scattering. The
expression (\ref{g3}) can thus be considered as a lower  bound to the
actual bare value  of $g_3$ in the Bechgaard salts \cite{Emery83}.

It is worth noting that the electron gas model is a  continuum model
which is  valid at low energy, so the connection with the parameters $U$
and $V$ of the lattice model can only be considered as  approximate. For
example, in the high-temperature range or for sufficiently large 
couplings  the curvature of the band may become relevant and  it is not
clear if a continuum theory will work well quantitatively
\cite{Mila95b,Mila96b}. The point at issue is  the `input' parameters 
 of the electron gas model. These are likely to enter  
 in the theory as   screened quantities due to non-logarithmic  many-body
effects and  they   
should then  differ from the  bare parameters of the  lattice model.

\subsection{Scaling theory}
A  simple feature of  logarithmic divergences like (\ref{log}) is the
lack  of  a particular scale in the energy interval between
$E_F$ and $T$. Another way to put it is to say that there is {\it
scale invariance} since each part of the integral in (\ref{log}),
whatever its size inside the interval, gives a logarithmic contribution
that is independent of either $T$ or $E_F$ \cite{Wilson75}.
  Scaling that is present at the lowest order will carry over not
only to leading, but also to  next-to-leading, etc., logarithmic
singularities of  the scattering amplitudes.
Therefore the  perturbation theory and in  its turn the
properties of the electron gas model as a whole will become
scale  invariant. 

The renormalization group method is an appropriate device  to sum up
perturbative series of this sort. Schematizing the basic idea behind this
procedure, it consists  of the successive partial integrations of  outer
energy-shell band electron states in the partition function $Z$ \cite{Bourbon91,Bourbon95,Solyom79}.
Consequent to this  reduction of  electronic degrees of freedom,
 the Hamiltonian  keeps the same
form except for the scaling or the renormalization of the coupling
constants. Other static and dynamical quantities that can
be worked out using $Z$ such as the response functions, the one-particle
spectral weight, the density of states, etc., will also show
renormalization. The recursion relation  of the hamiltonian
in
$Z$ is meant  by the following transformation
$$
Z =\ {\rm Tr}_<{\rm Tr}_{\rm o.s} \ e^{-\beta H} \propto \
{\rm Tr}_< \ e^{-\beta H_{d\ell}}\ldots
$$
\begin{equation}
  \ldots \propto \ {\rm Tr}_<{\rm Tr}_{\rm o.s} \ e^{-\beta
H_{\ell}}\propto {\rm Tr}_<\ e^{-\beta H_{\ell +d\ell}},
\end{equation}  
where Tr$_{\rm o.s}$ is the partial sum of diagonal matrix elements in a outer energy shell
$\dm E_0(\ell)d\ell$ on both sides of the Fermi level. Here  $d\ell$ is 
the infinitesimal generator
for the bandwidth reduction
$E_0(\ell)
\to E_0(\ell+d\ell)= E_0(\ell) e^{-d\ell}$ and $E_0(\ell)=2E_F
e^{-\ell}$ corresponds  to the effective or scaled   bandwidth at the
step $\ell$.  Under the renormalization group transformation  
$${\cal
R}_{d\ell}[H_\ell]= H_{\ell +d\ell},
$$
 the flow of 
coupling constants results at one-loop level from the Peierls and
Cooper  logarithmic   divergences. These  differ in sign and lead
to important cancellations (interference) in the renormalization flow
\cite{Dzyaloshinskii72}.  Thus, after all possible cancellations being
made, the flow is found to be governed by the following set of equations
$$
 {d\tilde{g}_1 \over d\ell} = 
-\tilde{g}^{2}_1 + N_1,
$$
$$
{d(2\tilde{g}_2 -\tilde{g}_1) \over d\ell} =  \ \tilde{g}^{2}_3 + 
N_2, \hskip 1.5 truecm
$$
\begin{equation}
\hskip 1.1truecm{d\tilde{g}_3 \over d\ell} =  \ \tilde{g}_3
(2\tilde{g}_2 -\tilde{g}_1) + N_3.
\label{flowcouplings}
\end{equation} 
Here we have defined  $\tilde{g}_1\equiv g_1/\pi v_\sigma $,
$(2\tilde{g}_2-\tilde{g}_1) \equiv (2g_2-g_1)/\pi v_\rho$, and $
\tilde{g_3}= g_3/\pi v_\rho $, by introducing the renormalization of the
velocities 
$$
v_{\rho,\sigma}= v_F[1 \pm g_4/(2\pi v_F)].
$$
The terms $N_{i=1,2,3} \sim {\cal O}(\tilde{g}^3) $ give the
higher order (two-loop) contributions. 

What  immediately  comes out of these equations at the one-loop level
is the fact that 
$g_1(\ell)$ is decoupled from   the set of couplings $(2g_2-g_1)(\ell)$ 
and 
$g_3(\ell)$. Reverting to the expression (\ref{hamiltonian}) for the 
Hamiltonian, this means that long-wavelength charge and spin 
correlations are decoupled, a key feature of a Luttinger liquid known
as  {\it spin-charge separation}.

Consider the charge part, the flow moves along  hyperbolas defined by
the equation
 (invariant) $(2\tilde{g}_2-\tilde{g}_1)^2-\tilde{g}_3^2 = C$ (Figure
\ref{flow}). Therefore for
$g_1-2g_2 < \mid g_3\mid$, it drives both 
$g_3$ and
$2g_2-g_1$ to strong coupling where a singularity develops  in the charge
sector at
$\ell_\rho\equiv \ln E_F/T_\rho$, corresponding to the temperature scale
\begin{equation}
T_\rho= E_F\  e^{-1/\sqrt{C}}.
\end{equation}
This signals the presence of a {\it gap} $\Delta_\rho \sim T_\rho$ in the
charge degrees of freedom, which has the same origin as the
Mott-Hubbard gap in the one-dimensional Hubbard model at
half-filling \cite{Lieb68}.
\begin{figure}[ htb] 
\epsfxsize0.95\hsize
\epsffile{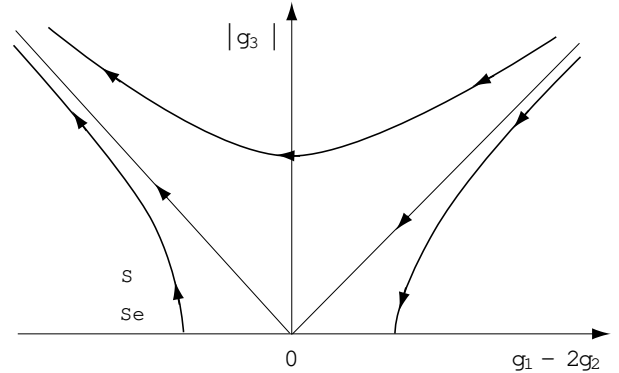}
\caption{Renormalization group flow of the couplings in the charge sector. The points S and Se
indicate locations of members of the \sulf \ and \sele \  series. }    
\label{flow}   
\end{figure}

As for the spin part it is governed by the flow of $g_1$ and at the 
one-loop level, one finds at once
\begin{equation}
g_1(\ell)= {g_1\over 1+ (\pi v_\sigma)^{-1} g_1 \ell}.
\label{spincoupling}
\end{equation}
The case of practical interest is a repulsive  $g_1
> 0$,   
where $g_1(\ell\to \infty) \to 0$, and which becomes  marginally
irrelevant. Thus referring  to  the expressions (\ref{couplings}) and (\ref{g3}), which
should hold for the Bechgaard salts and their sulfur analogs, the
domain  of interest for these systems  would be located  on the left-hand-side of the flow diagram
in Figure \ref{flow}. In this scenario,  the properties of 
\sele 
\ and \sulf,  should scale towards those of the purely
one-dimensional Hubbard model at half-filling which shows an insulating
state and gapless spins degrees of freedom.

At the
two-loop level, the recursion relations for the couplings   are obtained  using the expressions
\cite{Kimura75}:
$$
N_1= -\dm
\tilde{g}_1^3, \hskip 1.2 truecm
$$
$$
\hskip .2 truecm N_2= -
\dm
\tilde{g}_3^2(2\tilde{g}_2-\tilde{g}_1), 
$$ 
$$ 
\hskip 1.4 truecm N_3= -{1\over 4}\tilde{g}_3(2\tilde{g}_2-\tilde{g}_1)^2 - {1\over 4}
\tilde{g}^3_3.
$$
  Spin-charge separation is preserved
and the singularity at $T_\rho$, albeit removed for $g_3$ and
$2g_2-g_1$,  is replaced as $\ell \to \infty$ by a strong  coupling
fixed point at $g_3^* \to 2\pi v_\rho$
$g_2^*\to \pi v_\rho$ and $g_1^*\to 0$, which is still indicative 
 of a gap $\Delta_\rho$ in the charge. 

\paragraph{Scattering time and transport.$-$}

A rapid  growth of Umklapp scattering as $T_\rho$ is approached
from above carries with it dissipation of  momentum which  increases
resistivity. The two-loop evaluation of the imaginary part of the self-energy in the 
one-particle Green function leads to the following  expression for the 1D electron-electron
scattering rate  as a function of temperature \cite{Gorkov73}:
\begin{equation}
\tau^{-1}\sim [g_3(T)]^2 T.
\end{equation}
 The insulating behavior of \sulf\  and (TMDTDSF)$_2$X as the temperature is 
decreased  in the normal phase (Figure \ref{Resistivity}) has been ascribed  to the presence of
$T_\rho$ and to the relevance of  one-dimensional Umklapp scattering  in these materials
\cite{Emery82,Brazo85,Bourbon86}. As the temperature is lowered, however, the increase of $g_3$ is
faster than the phase space factor 
$1/T$   so that a metallic phase should be absent at all temperature which is not corroborated by
experiments (Figure \ref{Resistivity}). This result is confirmed by more elaborate calculations
\cite{Giamarchi91}. It was  pointed  out   that in systems like \sele \ and \sulf, 
$g_3$  is not the only source of Umklapp scattering \cite{Giamarchi97}. Actually
in the temperature range where $T> \Delta_D$, the influence of dimerization is small and the  
hole  band becomes essentially  3/4-filled  (or 1/4-filled for an electron band)
which introduces  an Umklapp term that is specific to this
order of commensurability. The quarter-filling Umklapp term noted $g_{1/4}$ will give rise to an
additional scattering  channel for the carriers which allows to  write the total 
electron-electron  scattering rate as the sum of two contributions:
\begin{equation}
\tau^{-1} = \tau^{-1}_{g_3} + \tau^{-1}_{g_{1/4}}.  
\end{equation}
Here $\tau^{-1}_{g_{1/4}}$ is expected to be the dominant contribution at high enough
temperature  and since  the growth of $g_{1/4}(T)$ is  less rapid than for $g_{3}(T)$ as one
moves down in temperature, this would give rise to a metallic behavior over a sizable domain of the
normal phase.  Although the 
$g_{1/4}$ term could be introduced in the above renormalization group approach, its influence  on
the  scattering rate will be rather discussed in more details  in the context of the bosonization
method.

\paragraph{Spin susceptibility.$-$}

The absence of any anomaly in the spin susceptibility near $T_\rho$ and its temperature variation
  can also be accounted for in the one-dimensional picture.
According to  (\ref{spincoupling}) and (\ref{hamiltonian}), the spins remain  gapless for
$g_1>0$  but the logarithmic screening of $g_1$  introduces  a
scale dependent  exchange coupling between left and right spin
densities, which in its turn,  affects the temperature variation of the
spin susceptibility. Since within the electron gas model, long wavelength
spin-density excitations  are  non-interacting modes, these can be
appropriately described within a  RPA type of calculation
\cite{Bourbon93}, which  yields
\begin{equation}
\chi_s(T) = {2\mu_B^2(\pi v_\sigma)^{-1}\over 1- (2\pi v_\sigma)^{-1}g_1(T)}.
\label{chis-oneD}
\end{equation}
 This expression for the susceptibility is of interest in many respects.
One first verifies that as $T\to0$, $g_1(T) \to 0$ and $\chi_s\to
2\mu_B^2(\pi v_\sigma)^{-1}$, a result that agrees with the exact result
of Shiba  for the Hubbard model at small $U$ \cite{Shiba72}. The
enhancement then differs from the  unrenormalized Hartree-Fock theory 
(\ref{FLchi}), for which  the logarithmic screening of $g_1$ is absent
and where $F^a = \dm N(E_F)(g_1 + g_4)
$ is  larger in amplitude.  Secondly, owing to the logarithmic screening
of the backward scattering, the  susceptibility shows a temperature
variation   that  can be made   congruent with    observation  made for
both
\sulf\  and
\sele \ series \cite{Wzietek93}. Take for example the
 case of (TMTSF)$_2$X; when  $\chi_s$ data are corrected for  thermal
dilatation and  restored to their  constant volume  values,  the
amplitude  of $\chi_s$ at 300 K  is found to be about 30\% larger than  at
50 K, below which it becomes essentially flat in temperature. As shown by
Wzietek
{\it et al.} \cite{Wzietek93},  by taking in the  Hubbard limit
$g_1=g_4= aU\sim 
\pi v_F$  and typical band calculation values
$E_F\sim 3000$K for the Fermi energy, quite reasonable agreement can be
obtained   between the predicted and the observed temperature variations
of $\chi_s$ in this temperature interval. The same type of analysis
carries over to (TMTTF)$_2$X compounds \cite{Maaroufi83,Parkin83}, where
similar figures for the electron coupling constants and the use of
somewhat lower values of 
$E_F$ lead to similar agreement \cite{Wzietek93}.  In this frame then,
the temperature variation of the susceptibility in the normal phase can
provide useful information on the range of  one-dimensional physics in
these materials.

\paragraph{One-particle spectral properties.$-$} 
The two-loop corrections are also of interest 
for the quasi-particle weight $z$ (cf. Eq. \ref{GreenFL}) at the Fermi level which is governed by 
the scaling equation
\begin{equation}
{d\over d\ell}\ln z = -{1\over 16}[C + {3\over4}(\tilde{g}_1^2
+\tilde{g}_3^2) ].
\end{equation}
The integration leads to the  power law decay  
\begin{equation}
z(x) \sim x^{\alpha},
\end{equation}
where  $x$ can be identified either with $\omega$ (for  $T=0$)  or $T$ (for 
$\omega< T$). As for the power law exponent  $\alpha $, it  is   positive and in general
non-universal indicating that at zero temperature there are no quasi-particle states at the Fermi
level, a  characteristic feature of the Luttinger liquid.  The spectral function is 
\begin{equation}
{\rm Im}\ G_p(k_F^0,\omega) \sim \ \mid \omega\mid^{\alpha-1},
\end{equation}
which also varies as a power of the energy. 

Attempts to interpret ARPES data of (TMTSF)$_2$X in the normal metallic
phase by using a   power law behavior of this form for the spectral weight
have shown to  require rather large values of
$\alpha$ \cite{Dardel93}, which are not accessible through the above 
perturbative  renormalization group method.  Away from the strong
coupling fixed point, the present approach predicts
that 
$\alpha
$ is a rather small quantity in the metallic phase. More elaborate calculations like the
bosonization technique (see below) and numerical calculations allow to give a more precise 
treatment of the exponents of  spectral functions at low energy
\cite{Voit95,Schulz90,Mila93}.

\paragraph{Response functions.$-$} 
The elementary Cooper and Peierls logarithmic divergences (\ref{log}) of
the interacting electron gas are also present order by order   in
the perturbation theory of  response functions in the
$2k_F^0$ density-wave  and superconducting  channels. A  scaling
procedure can thus be applied in order to obtain the asymptotic
properties of the real part of the retarded response functions which  we
will note 
$ \chi_\mu(\ell) $. It is convenient to introduce auxiliary response
functions noted
$\bar{\chi}_\mu(\ell)$ \cite{Dzyaloshinskii72}, which are defined 
\begin{equation}
\chi_\mu(\ell) = -(\pi v_F)^{-1} \int_0^\ell \bar{\chi}_\mu(\ell') \ d\ell'.
\label{response}
\end{equation} 
In the presence of Umklapp scattering,  the following   susceptibilities
will be considered: for the Peierls channel at 2$k_F^0$, one has the
`on-site' $\mu= $ SDW for  spin-density-wave  and $\mu=$ BOW  for the  `bond-order-wave'; in the 
Cooper channel, one has    singlet $\mu=$ SS and triplet  $\mu=$ TS
superconducting correlations. At two-loop level, the auxiliary responses
are governed by the flow equations
\begin{equation}
{d\over d\ell}\ln\bar{\chi}_\mu = \ g_\mu(\ell) + \dm [g_1^2(\ell) + g_1^2(\ell)
-g_1(\ell)g_2(\ell) + \dm g_3^2(\ell)],
\label{flowresponse}
\end{equation}
where   $g_{{\rm SDW}}= g_2(\ell) +
g_3(\ell)$  and  $ \ g_{{\rm BOW}}= g_2(\ell) + g_3(\ell)-2g_1(\ell)$) for  the Peierls channel; and
$g_{\rm SS} = -g_1(\ell)-g_2(\ell)$, and  $g_{\rm TS} = -g_2(\ell)+
g_1(\ell)$  for  the Cooper channel. In the repulsive sector which is of
 interest for the Bechgaard salts, only SDW and BOW correlations
develop singular responses\footnote{Here we do not consider the 4$k_F^0$ response of the electron
gas which is also singular in this sector \cite{Tutis90}.}  at
$2k_F^0$ (Figure
\ref{Diag1D}). Using (\ref{flowcouplings}), the asymptotic behavior of the auxiliary response
as a function of temperature  is found to be 
\begin{equation}
\bar{\chi}_\mu(T)\approx X_\mu(E_F/\Delta_\rho)\left({\Delta_\rho\over T}\right)^{-\gamma^*_\mu},
\end{equation}
which varies as a power of the temperature. In the presence of a Mott-Hubbard gap, the exponent for
both bond-order-wave and site spin-density-wave $\gamma^*_{{\rm
SDW}}=\gamma^*_{{\rm BOW}}=3/2$ are then of the order of unity. Here $X_\mu $ are 
scaling factors that   give the power law behavior in the weak
coupling domain, where $g_3$ is weak and $\gamma_\mu$ is smaller than
unity. The possible singularities of the electron gas at half-filling are
summarized in  Figure \ref{Diag1D}.  
\begin{figure}[ htb] 
\epsfxsize0.95\hsize
\epsffile{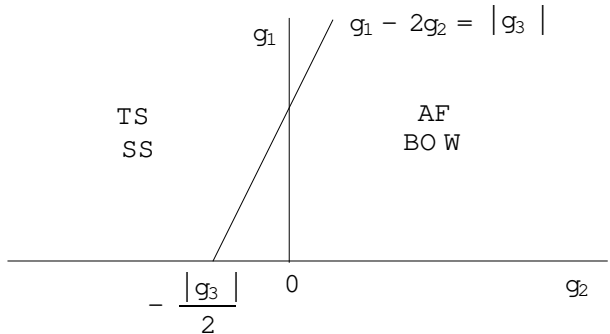}
\caption{Phase diagram of the electron gas model in the repulsive sector ($g_1 >0$) and in
presence of Umklapp scattering.   }    
\label{Diag1D}   
\end{figure}

These results of the one-dimensional theory are
of interest since they indicate that the relevance of Umklapp
processes at half-filling eliminates the possibility of 
charge-density-wave  correlations which  usually drive an incommensurate
system to  
 a Peierls instability $-$ as it is the case for  TTF-TCNQ
\cite{Emery82}. Instead of a Peierls instability, 
antiferromagnetic correlations  dominate.   This is consistent
with  the fact that antiferromagnetism plays a very important part of the
phase diagram of (TM)$_2$X (Figure \ref{Diag-phase}).   Bond-order-wave
correlations, however, subsist according to the renormalization group
results  $-$ 
 on almost equal footing with antiferromagnetism. For  dimerized  TMTTF 
or  TMTSF stacks, these would correspond to the correlation  of the
carrier tunneling across the `bond'  separating each  dimer at wave vector
$2k_F^0$. When the coupling of these electronic correlations to acoustic phonons is suitable, the
system has the possibility to develop a  structural instability called a spin-Peierls instability,
which is the analog of the Peierls distortion  when the  electronic part of
the system is Mott-Hubbard insulating instead of metallic away from half-filling.

One-dimensional precursors to a spin-Peierls instability  has been
clearly identified by X-ray diffuse scattering experiments  in (TMTTF)$_2$PF$_6$,
(TMTTF)$_2$AsF$_6$ and (TMDTDSF)$_2$PF$_6$, well above the transition
temperature and for low pressure conditions (Figure \ref{Xray}). 
As one moves to the right-hand-side of
the phase diagram in Figure \ref{Diag-phase}, these precursor effects of the normal phase dwindle in
amplitude. In (TMTSF)$_2$PF$_6$ for example, where the normal phase is
metallic and Umklapp processes are less singular, $2k_F^0$ X-ray diffuse
scattering   is  small in amplitude and shows a slow increase as a
function of temperature down to 50K,  below which it starts  
decreasing \cite{Pouget82,Pouget96,Pouget97}. These X-ray features convey in a way the
strength of the  one-dimensional bond-order-wave singularity  in the
electronic system, which turns out to be consistent with the prediction
of the one-dimensional theory. Indeed,    the explicit evaluation of
the  amplitude of the scaling factor
$X_{{\rm BOW}}(T)\ll X_{ {\rm SDW}}(T)$ $-$ in the absence of a charge
gap $-$ is found to be much  smaller for BOW  than for  antiferromagnetism.
It is worth emphasizing that although such lattice precursors are weak
in amplitude  in systems like
\sele, they have the same one-dimensional character  as in  
(TMTTF)$_2$PF$_6$, or in Peierls systems like  TTF-TCNQ \cite{Pouget76} and KCP
\cite{Comes73}.  Therefore  the temperature profile of the X-ray diffuse scattering
intensity (Figure \ref{Xray}) can be seen as a precious indication about the extent to which
one-dimensional physics is relevant in the normal phase of these
systems. 
\begin{figure}[ htb] 
\epsfxsize0.95\hsize
\epsffile{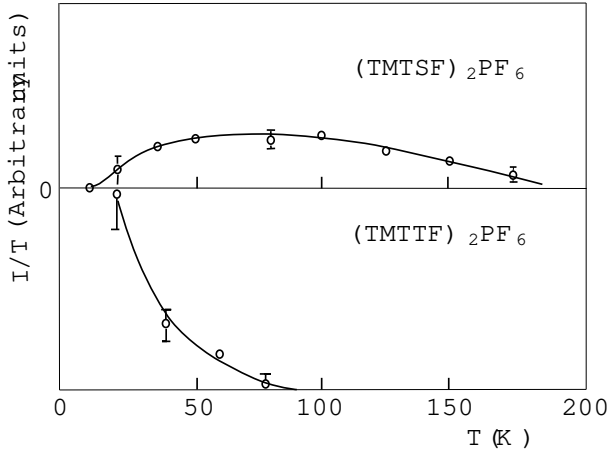}
\caption{Temperature dependence of the $2k_F^0$ scattering intensities for (TMTSF)$_2$PF$_6$ (top)
and (TMTTF)$_2$PF$_6$ (bottom) in the normal phase, after \cite{Pouget96}.}    
\label{Xray}   
\end{figure}

\subsection{The concept of Luttinger liquid and bosonization}
In the presence of strong coupling  for the charge couplings  $g_3$ and
$2g_2-g_1$, the renormalization group approach, which relies on the
perturbation theory of original fermions,    becomes less accurate  in
 obtaining quantities like the power law indexes. For more
quantitative results, it is therefore suitable to call on  a different
approach   which can even allow in some  special cases an exact
solution of the problem.  In what follows we would like to  give a brief
presentation of the relevant results that can be obtained by
the bosonization approach and how  this  supplementary  amount of
information can help to go deeper in our understanding of the normal
phase of the Bechgaard salts and their sulfur analogs. 

 A feature of peculiar importance  in one dimension is that long wavelength
 charge or spin-density-wave oscillations constructed by the combination
of electron-hole pair excitations at low energy form  extremely stable
excitations \cite{Voit95}. These actually have no available phase space to decay $-$ in
contrast to the situation found in a Fermi liquid where a damping
exists \cite{Pines66}.   Quasi-particles are absent at low energy for a one-dimensional
system of interacting electrons  and are replaced by collective  
acoustic excitations for both spin and charge degrees and freedom which
turn out to be the true eigenstates of the system \cite{Voit95}. Following
the example  of Debye sound waves in a one-dimensional solid, their
dispersion relations are linear and reads
\begin{equation}
\omega_{\sigma,\rho}= u_{\sigma,\rho}\mid q\mid.
\end{equation}
where $u_{\sigma,\rho}$ is the velocity of each of these modes.   Collective 
oscillations of the electron gas will then obey boson statistics and the connection between 
the original Fermi  and  boson fields  is made possible  through the following relation 
$$
\psi_{p,\sigma}(x)  =  L^{-\dm} \sum_k a_{p,k,\sigma}\ e^{ikx}
$$
\begin{equation}
\sim \lim_{\alpha_0 \to 0} {e^{ipk^0_Fx}\over \sqrt{2\pi\alpha_0}} \exp\Bigl(-{i\over
\sqrt{2}}[p(\phi_\rho + \sigma \phi_\sigma) + (\theta_\rho + \sigma\theta_\sigma)]\Bigr),
\end{equation}
where the spin and charge
phase fields $\phi_{\rho,\sigma}$  have been introduced \cite{Schulz98}. These satisfy the
commutation relation
\begin{equation}
[\Pi_{\nu^\prime}(x^\prime),\phi_\nu(x)]= -i\delta_{\nu\nu'} \delta(x-x'),
\end{equation}
where   $\Pi_{\nu}(x)$ is   the momentum conjugate to
$\phi_\nu(x)$ and is defined by $\theta_\nu(x)= \pi\int \Pi_\nu(x') dx'$.
Following   Schulz {\it et al.}~\cite{Schulz98}, the phase variable
representation 
  allows to rewrite the  full  electron gas Hamiltonian in the form
$$
H= \sum_{\nu =\rho,\sigma}\dm \int  \left[ \pi u_\nu K_\nu \Pi_\nu^2 + u_\nu (\pi K_\nu)^{-1}
\left({\partial
\phi_\nu \over \partial x}\right)^2 \right] dx 
$$
$$
 + \  {2g_1\over (2\pi\alpha_0)^2} \int  \cos(\sqrt{8}\phi_\sigma) \ dx
$$
$$
+ \  {2g_3 \over (2\pi\alpha_0)^2}\int  \cos(\sqrt{8}\phi_\rho) \ dx
$$
\begin{equation}
\ + \  {2g_{1/4} \over (2\pi\alpha_0)^2}\int  \cos(2\sqrt{8}\phi_\rho) \
dx.
\label{hamiltboson}
\end{equation}
The harmonic part of the phase Hamiltonian corresponds to the Tomanaga-Luttinger model with no
backscattering and Umklapp terms. It is exactly solvable, the spectrum
shows  only collective excitations and all the properties of the model
then become entirely governed by the velocity $u_\nu
$ and the `stiffness constant'
$K_\nu$ of  acoustic  excitations,
which  are functions of the
microscopic coupling constants and differ for the charge and spin. The
collective modes for the spin and charge  are decoupled and this leads  to spin-charge separation of
the  Luttinger liquid. When backscattering or/and Umklapp scattering are
non zero, the  sine-Gordon terms in (\ref{hamiltboson}) do not allow an
exact solution in general $-$ except for particular  values of couplings
corresponding to the  Luther-Emery model
\cite{LutherEmery74,Emery79}. Note that in (\ref{hamiltboson}), the contribution coming
from quarter-filling Umklapp processes has been added to the phase Hamiltonian \cite{Giamarchi97}.
This source of Umklapp scattering should dominate in the  temperature domain where $T\gg \Delta_D$
and the influence of dimerization gap should be small. In the Hubbard limit, its  bare amplitude  
is given by 
\begin{equation}
g_{1/4}\approx aE_0 \left({U\over E_0}\right)^3.
\end{equation}
Renormalization group  method can be used to determine the flow of
renormalization for  $(g_1,g_3,g_{1/4})$ when these act as perturbations for the  Luttinger liquid
parameters
$(K_\nu,u_\nu)$ as a function of energy.

 The  properties of the  model in the repulsive sector can be obtained  at low energy  by
looking  close to the fixed point. For  a rotationally invariant  system at
half-filling or quarter-filling, one gets $g_1^*\to 0$, while 
$g^*_3(g_{1/4}^*)$ and $2g_2^*-g_1^*$ reach strong coupling. In
this  case, one has
$$
K^*_\sigma \to 1, \ \ \   u^*_\sigma \to v_\sigma 
$$
\begin{equation}
K^*_\rho  \to 0, \ \ \   u^*_\rho \to (2\pi)^{-1}\sqrt{ (2\pi
v_\rho)^2 - (2g^*_2-g_1^*)^2},
\label{fixedpoint}
\end{equation}
which implies a vanishing velocity for collective charge excitations and a gap given by  
\begin{equation}
\Delta_\rho \sim E_0\left(g_{U}\over E_0\right)^{1/ [2(1-n^2K_\rho)]},
\end{equation}
where $n=1$ for $g_U=g_3$, and $n=2$ for $g_U=g_{1/4}$ at half-filling and quarter-filling, 
respectively. It should be stressed here that in   the pure quarter-filled (resp. half-filled)
case,  the existence of a charge gap is subjected to the condition $K_\rho < .25$ (resp. $K_\rho
<1$) $-$ here $K_\rho=1 $ refers to free electrons \cite{Mila93}.     In the metallic
range where
$g_3$ and $2g_2-g_1$ are not too large, $K_\rho$ and $u_\rho$ are
finite,  while  one can  take $K_\sigma=1$ for the spin part if the
system is  invariant under rotation. The Luttinger liquid
parameters 
$u_\rho$, $K_\rho$ and $u_\sigma$  at low energy can also be obtained using numerical
methods applied to lattice models at different band fillings
\cite{Schulz90,Mila93}.

Given the values of the Luttinger liquid parameters, we can now look at the properties of the
system  using the harmonic part of the Hamiltonian. Thus the spin-spin
correlation function at $2k_F^0$ can be expressed as
statistical averages over phase variables, which can be explicitly evaluated
\cite{Schulz98,Voit95}. At equal-time for example, one gets $-$ after dropping logarithmic
corrections $-$  the power law decay   
$$
\chi(x) =  \langle {\bf S}(x)\cdot {\bf S}(0)\rangle 
$$
\begin{equation}
          \ \ \ \ \ \ \sim \ {\cos(2k_F^0x)\over x^{1+K_\rho}}. 
\end{equation}     
The stiffness constant $K_\rho$ then entirely governs the
algebraic decay of the antiferromagnetic correlations. The spin response at $2k_F^0$  is given by
\begin{equation}
\chi(2k_F^0,T) \sim T^{-1+K_\rho}.
\end{equation}
 The power law singularity is therefore
the strongest in the presence of a charge-degrees-of-freedom gap where
$K_\rho=0$, which  actually corresponds to the Heisenberg universality class.
	 
	\paragraph{Nuclear spin-lattice relaxation: insulating domain.$-$} As mentioned earlier a good
insight into spin correlations can be obtained through  nuclear spin-lattice relaxation.
According to (\ref{Moriya}), $T_1^{-1}$ is related to the imaginary part of
the dynamical spin response. In a Luttinger liquid  ${\rm
Im}\chi(q,\omega\to 0)$ is strongly peaked at $q=0$  and $2k_F^0$, so
the integration over all modes  makes both types of low-lying
excitations  contributing to the relaxation rate \cite{Bourbon89,Bourbon93}. 

Near $q=0$, on finds for the spectral weight of spin excitations:
\begin{equation}
{\rm Im}\chi(q,\omega)=  {1 \over 4 v_\sigma} \sum_p{pv_\sigma
q\delta(\omega- pv_\sigma q)\over [1 - (2\pi v_\sigma)^{-1}g_1(T)]^2}.
\end{equation}
The presence of a delta function in this expression indicates the
absence of damping for paramagnons  in a Luttinger liquid $-$ in
contrast to a Fermi liquid (see Eq.~\ref{spinspectral}) $-$ this also indicates that spin 
collective excitations are  exact eigenstates of the system in one
dimension (similar conclusions  also hold for the
charge part). 
Now close to $q=2k_F^0$, one can use the following expression \cite{Bourbon91}: 
\begin{equation}
{\rm Im }\chi(q\sim 2k_F^0,\omega\to 0) \sim T^{-1+K_\rho}
\ \omega,
\end{equation}
which is valid in the low frequency limit. The summation over all
q-vectors in the expression of the nuclear relaxation rate can then be evaluated at once and yields
\cite{Bourbon93,Bourbon89} 
\begin{equation}
T_{1}^{-1} = C_{0}T\chi_s^{2}(T) +
C_{1}T^{K_\rho},
\label{relax}
\end{equation}
where $C_0$ and $C_1$ are $T$-independent parameters.
For a Luttinger liquid, the nuclear relaxation rate is thus enhanced with
respect to the Korringa law. The temperature variation of
the enhancement  then combines 
 two different contributions which leads to a
characteristic temperature profile for the relaxation rate.  In the
high temperature domain, antiferromagnetic correlations are weak and
the relaxation rate  is dominated by the enhancement of the square of the
spin susceptibility, which leads according to (\ref{chis-oneD}) to an upward curvature of $T_1^{-1}$
as a function of $T$, while  it is the other way
round in the low temperature domain (see Figure \ref{T1SeClO4}); there, the contribution of
paramagnons becomes smaller to the benefit of antiferromagnetic spin
correlations which eventually merge as the
dominant source of enhancement which varies as power of the temperature. 

As previously mentioned, the  exponent  $K_{\rho}$ is
zero for a 1-D Mott-Hubbard insulator and antiferromagnetic spin fluctuations contribute a constant
term to the relaxation rate. A canonical example for this behavior is observed
in
(TMTTF)$_{2}$PF$_{6}$ \cite{Wzietek93}, as  $T_{1}^{-1}$ plotted versus
the measured value of 
$T \chi_{S}^{2}(T)$  shows a finite intercept at $T = 0$ (Fig. \ref{T1chisT2}). Other
illustrations for the viability of the 1-D model for spins are also given
by systems such as (TMTTF)$_{2}$Br
\cite{Wzietek93} and
(TMDTDSF)$_{2}$PF$_{6}$ \cite{Gotschy92}, exhibiting a less pronounced
Mott insulating character with $T_\rho$ only in the 80 $- $ 200~K
range.
$T_{1}^{-1}$ follows a linear law  without any finite  intercept at low
temperature as long as $T > T_\rho$ \cite{Jerome94}. This is the expected
behavior  when the $q=0$ fluctuations are predominant.
Antiferromagnetic  fluctuations show a much stronger singularity below
$T_\rho$ where $K_\rho= 0$ and contribute according to (\ref{relax}) an
additional constant term to the relaxation  at very low temperature
(Fig.~\ref{T1chisT2}).
\begin{figure}[ htb] 
\epsfxsize0.95\hsize
\epsffile{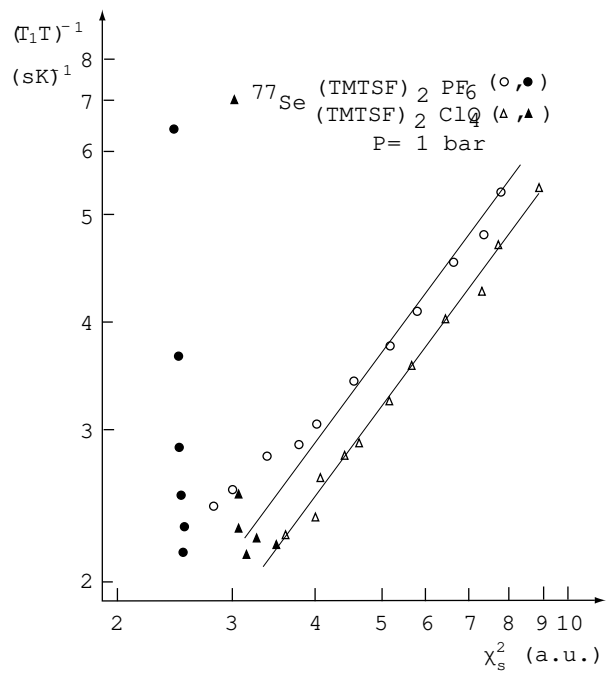}
\caption{$^{13}$C $T_1^{-1}$ vs measured EPR $T\chi_s^2$ data for (TMTTF)$_2$PF$_6$ and
(TMTTF)$_2$Br (top); $^{77}$Se $T_1^{-1}$ vs measured Faraday
$T\chi_s^2$ data for (TMTSF)$_2$PF$_6$ (bottom), after \cite{Bourbon89,Wzietek93}.}    
\label{T1chisT2}   
\end{figure}

	\paragraph{Antiferromagnetism ordering in the presence of a charge gap.$-$}
The intrastack charge
localization resulting from the effect of Umklapp scattering is also accompanied by a confinement
of the carriers on individual stacks. This confinement can be viewed as a result of a
reduction in the transverse kinetic coupling $t_{\perp b} $ due to the
correlations in the 1-D Mott state (i.e. the renormalization of the
transverse crossover $T_{x^1}$ in one-particle motion discussed at the beginning
of the section 3 on the Fermi liquid) to the extent of becoming completely
irrelevant \cite{Bourbon86,Bourbon95}.   In spite of the absence of a cross-over in the coherent
single-particle motion, the existence of long-range antiferromagnetic
order, which is known  to be found in the range of 5$-$25K  in \sulf, 
indicates that spin correlations end in deconfinement at sufficiently low
temperature. Following the example of localized spins in the Heisenberg
limit, this is made possible through a kinetic  exchange interaction
$J_{\perp b}$ between spins of neighboring chains \cite{Bourbon86,Brazo85,Bourbon95}, which can also
be viewed as the interchain transfer
 of bound electron-hole pairs. In Mott-Hubbard systems like
\sulf, however, the  localization of carriers is less pronounced than in
the Heisenberg limit and virtual processes take place  over larger
distances
$\xi_\rho \sim v_F/\Delta_\rho
\gg a$,  which magnifies the exchange to  yield 
$$
J_{\perp,b} \approx {\xi_\rho\over a}
{t_{\perp b}^{*2}\over \Delta_\rho},
$$
where $t_{\perp b}^*$ is the effective but finite interchain hopping taking place at  the
energy scale  $\Delta_\rho$. Interchain exchange is the key mechanism for
promoting   antiferromagnetic long range order in \sulf\  over a
sizable range of pressure, and to a lesser extent
in \sele.    By its coupling to singular antiferromagnetic fluctuations along the chains, the
condition for the onset of critical ordering is given by $J_{\perp,b} \chi(2k_F^0,T_N) \sim 1$,
which yields
$$
 T_N \sim {t_{\perp b}^{*2}\over \Delta_\rho}.
$$
  This leads to a characteristic  increase of 
$T_N$ as a function of pressure which is observed experimentally \cite{Klemme95,Brown97}. This 
lends  additional support for  the existence  of a  Luttinger liquid
state  severed of its charge component in the normal phase of these compounds (see Fig.
\ref{Diag-phase}).

\paragraph{Nuclear spin-lattice relaxation: conducting regime.$-$}
As for the spin sector in the
conducting regime, an approach to the 1-D exponent $K_\rho$ can be attempted looking at the
spin-lattice relaxation rate for which the $2k^0_F$ contribution should become predominant at low
temperature where  $0<K_{\rho} <1$ and a
deviation in the law $1/T_1 \sim  
 T\chi_{S}^{2}(T)$ can  be anticipated. This is precisely what is
observed from the $^{77}$Se-NMR in
(TMTSF)$_2$PF$_6$, where the log-log plot of $(T_1 T)^{-1}$ versus
$\chi_{s}^{2}(T)$ shows a deviation from linearity below $150K$ or so (Fig.~\ref{T1Tchis2}),
signaling the onset of $2k_F^0$ fluctuations. These AF fluctuations become even more visible when
$T_1^{-1}$ is plotted versus
$T$  for (TMTSF)$_2$ClO$_4$  as shown in Fig.~\ref{T1SeClO4}  $-$  or for (TMTSF)$_2$PF$_6$ under a
pressure which is large enough to suppress the SDW ground state \cite{Wzietek93}. A
temperature independent contribution arises below $50-30$~K (i.e., much above the temperature regime
for 3-D critical fluctuations). The non-critical enhancement of $T_1^{-1}$ is thus too large to be
ascribed to a Fermi liquid contribution which should behave linearly in a
temperature regime where the spin susceptibility is constant (see eq.~(\ref{relaxuniform})). The
comparison between the behavior of (TMTSF)$_2$PF$_6$ at low temperature and compounds such as
(TMTTF)$_2$Br or
(TMDTDSF)$_2$PF$_6$ which evolve towards the Mott insulator limit at low
temperature shows that the strong coupling limit is not reached in case
of selenium compounds, namely $K_{\rho}$ although renormalized in
temperature \cite{Bourbon87,Wzietek93}, never reaches the limit
$K_{\rho} =0$. A value $K_{\rho}\approx 0.1-0.2$ would be consistent with ambient pressure NMR data
in (TMTSF)$_2$PF$_6$ at $T\gaeq 50$K. Similar figures for  $K_\rho$ have been suggested in
(TMTSF)$_2$ClO$_4$ (or (TMTSF)$_2$PF$_6$ under pressure) \cite{Bourbon84,Creuzet87b}, although the
influence of interchain coupling in that region is probably not small and should contribute to some
extent to the relaxation rate.

	As shown in Fig.~\ref{T1SeClO4}, another regime of relaxation is reached at
very low temperature $(T < 8$ K), where a behavior $1/T_1 \propto T$ is
recovered \cite{Bourbon84,Takahashi84,Takigawa86,Azevedo81}. The low temperature
regime looks like a Korringa law with an enhancement factor of the order
10 with respect to the regime
$T > 30 $~K. It has been first proposed that  this 
change in behavior for the enhancement originates in the dimensionality
crossover  of one-particle coherence and the restoration of a Fermi liquid component in two 
 directions. 
	It is still an open problem to decide whether the Fermi liquid
properties recovered below 8~K are those of a
2-D or 3-D electron gas. Furthermore the intermediate temperature
regime 8~K $ \laeq\  T \ \laeq $ 80~K requires an improvement of the theory taking
into account all transient effects due to interchain coupling.

\paragraph{	The Wilson ratio in a Luttinger liquid.$-$ }
The  Fermi liquid which is recovered below 8~K is not
following the canonical behavior as shown by its response to a magnetic
field. According to high-field NMR studies of $^{77}$Se in
(TMTSF)$_2$ClO$_4$ \cite{Caretta95}, the uniform spin susceptibility $-$ as obtained from the Knight
shift
$-$ is found to be insensitive to the application of a magnetic field up to
15~T along $c^{\star}$. As already  discussed in Section~3, the Sommerfeld constant of the
electronic specific heat shows an important field dependence \cite{Brusetti83},   which cannot be
explained with the Fermi liquid model (Figure~\ref{gamma(H)}). The Luttinger liquid scenario,
however, offers an interesting avenue to understand these features if one considers this
experimental finding as reminiscent of the expected behavior in a slightly doped Mott insulator as
the metal-insulator transition is approached (i.e. band filling approaching half-filling). The
Wilson ratio in a Luttinger liquid reads \cite{Schulz91} :
$$
R_W = 
\frac{2v_{\rho}}{v_{\rho}+v_{\sigma}},
$$
in terms of charge and spin
velocities. Therefore, $R_W$ decreases as the Mott insulating regime is
approached since according to (\ref{fixedpoint}), $v_{\rho}\rightarrow 0$, while the spin
part $-$ which is independent of the Umklapp scattering $-$ is not sensitive to the
proximity of the Mott insulator. The behavior of the Wilson ratio under
magnetic field provides another example of marginal Fermi liquid
character in  2-D anisotropic conductor at low temperature and suggests 
that the spin  and charge separate as a result  the electronic confinement induced by a
magnetic field \cite{Behnia95}.

\paragraph{One-particle spectral properties of a Luttinger liquid.$-$}

The absence of quasi-particles in a Luttinger liquid is also manifest in  the one-particle spectral
properties.  As we have seen in the framework of the 1D renormalization group method in the gapless
case, there is a power-law decay of the one-particle spectral weight at the Fermi level. This
power law behavior is also confirmed in the framework of the bosonization technique
\cite{Voit93,Schonhammer92}, namely
\begin{equation}
A_p(k_F^0,\omega) \sim \ \mid \omega\mid ^{\alpha-1},
\end{equation}
where $\alpha= {1\over 4}(K_\rho +1/K_\rho -2) $. A  power law is also found for the
density of states 
$$
N(\omega) = \sum_p\int  A_p(k_F^0,\omega) \ dk
$$
\begin{equation}
        \sim \ \mid \omega \mid^\alpha.
\end{equation}
Thus as soon as right and left-moving carriers interact with each other, $K_\rho < 1 $ so that  
$\alpha > 0$, which leads to a dip  in the density of states  at the Fermi level.
As a function of temperature, one can replace $\omega$  by  $T$ and the density of states
vanishes as a power of the temperature. 

When Umklapp processes are relevant, $K_\rho $   decreases and $\alpha$ increases at low energy
leading to a pronounced depression of the spectral weight near $k_F^0$. Within a one-dimensional
scenario  for  the normal phase in the Bechgaard salts, this situation  is of
practical interest.  Experimentally both ARPES \cite{Grioni97} and integrated
photoemission \cite{Dardel93} signals fail  to detect any quasi-particle states in systems like 
(TMTSF)$_2$ClO$_4$ and (TMTSF)$_2$PF$_6$ (Figure \ref{photoemission}); instead, a quite pronounced
reduction of the spectral weight is found,  which one would be tempted to describe using the 
above expressions with a sizable value of $\alpha \simeq 1.5\ (K_\rho \simeq 1/8) $; a value which
we may add congruent with those extracted from a low-dimensional description  of   NMR
\cite{Wzietek93}, optical data \cite{Degiorgi98} (see below), and DC transport \cite{Moser98}.
Although photoemission results call into question the viability of the
Fermi liquid description in these systems, they
  confront us, however, with many difficulties that do not conform with the prediction of the
one-dimensional theory. Among them, the absence of dispersing structure and a power-law frequency
dependence that is spreaded out over  a large  energy scale of the order of 1~eV, may indicate  that
surface effects should be taken into account in the interpretation.

 ARPES experiments have also  been carried out in the insulating sulfur compound
(TMTTF)$_2$PF$_6$ (Figure \ref{photoemission}) and as expected \cite{Voit98}, the  charge gap is
clearly seen as an energy shift of
$\sim 100$~meV in the spectral weight,  a value that agrees with DC and optical transport.

\paragraph{Transverse and optical  transport.$-$} 
Transport properties along directions transverse to the stacking axis
can add to the understanding of this strange behavior \cite{Moser98}. If we consider the
$c$ direction which corresponds to the direction of weakest overlap, not
much optical work has been devoted to this direction but the absence of
any significant reflectance in the FIR regime for (TMTSF)$_{2}$AsF$_6$
at
$T = 30$~K has been attributed to the absence of coherent band transport
along this direction down to (at least) 30~K in fair agreement with band
calculations leading to $t_{\perp c} \approx 10-20$~K .

	Extending the arguments proposed for the incoherent transverse transport developed in
Q-1-D conductors to the 2-D conductors, one can infer as previously pointed out in Section 3 that
$\sigma_c$  is directly related to the physics of the $a-b$ planes and therefore could probe
whether transport proceeds via collective modes or independent quasi-particles.  The existence of  
maximum in $\rho_c$ at  
$T_{m}$ which evolves under pressure and reaches about 300~K at 10~kbar and the striking
different temperature dependencies for the {\it in} and {\it out} of plane resistances suggest an
interpretation in terms of a non-Fermi liquid approach at $T > T_{m}$. Assuming the $a-b$ planes can
be described by an array of chains forming 1-D Luttinger liquids at $T > T_{m}$ the
interplane transport can only proceed via the hopping of individual
particles. For such a situation to occur a particle has to be rebuilt out
of the Luttinger liquid by recombining its charge and spin components.
The particle can then hop onto a neighboring stack, contributing to
$\sigma_c$  and then decay into the Luttinger liquid again. The
transverse interplane conductivity has been derived theoretically when
the physics of electrons in chains is governed by a 1-D Luttinger regime
and becomes \cite{Moser98}:
\begin{equation}
\sigma_{c}(T)\approx t_{\perp c}^{2}\frac{e^2}{\hbar}
\frac{ac}{v_c^2}\left(\frac{T}{v_c}\right)^{2\alpha},
\end{equation}
where the exponent $\alpha$ enters the
density of states near the Fermi energy which is proportional to  $\mid\omega\mid^\alpha$,
as we have seen above. Following this model, $\rho_c$  should behave like $\rho_c(T) \approx
T^{-2\alpha}$  in the Luttinger liquid domain with $\alpha$ positive and ranging from infinity in a
Mott insulator to zero in a non-interacting Fermi liquid.

If one reverts to the constant volume data of Figure \ref{resistance-ac}, a law such as
$\rho_{c}(T)\approx T^{-1.4}$ fits the data fairly well above $T_{m}$,
i.e. $\alpha = 0.7$. Below 60 K,  there is hardly any difference between
constant $T$ and constant $P$ variations since the thermal expansion is
there greatly  suppressed.
As for the longitudinal
transport, the result is displayed in Fig. \ref{ResistV}, where a cross-over from a
superlinear to a linear (or sublinear) power law temperature dependence
is observed in the vicinity of $80$~ K. Fig. \ref{resistance-ac} emphasizes the
remarkable feature of (TMTSF)$_{2}$PF$_6$, namely, opposite temperature
dependencies for interplane and chain resistivities above $T_{m}$.

We shall see below that the transport properties above $T_{m}$ are
consistent with the picture of Luttinger chains in $(a-b)$ planes. The
exponent $\alpha = 0.7$ derived from the constant volume transverse data
leads to $K_{\rho} = 0.22$. This value of $K_{\rho}$ allows in turn a
prediction for the constant volume $T$-dependence for $\rho_a$. The only
scattering process through which electron-electron collisions can
contribute to resistivity in this 1-D electron gas occurs when the total
momentum transfer is commensurate with a Brillouin zone wave vector. For
the situation of a 1/4-filled 1-D band which is likely to apply to
(TMTSF)$_{2}$PF$_6$ as the dimerization can be forgotten in first
approximation  ($\Delta_{D}<t_{\perp b}$) , the resistivity due to
1/4-filled Umklapp becomes \cite{Giamarchi97}:
\begin{equation}
\rho_{a}(T) \approx g_{1/4}^{2} T^{16K_{\rho}-3}.
\end{equation} 
  A value
$K_{\rho} = 0.22$ would in turn lead to $\rho_{a}(T) \approx T^{0.52}$ 
 which is in
qualitative agreement with the sublinear $T$-dependence in Fig.~\ref{ResistV},
given the degree of arbitrariness which underlies the conversion
procedure.

	The temperature $T_{m}$ signals the beginning of a cross-over between a
high temperature Luttinger regime for the chains along $a$ and a 2-D
regime where the system leaves progressively the LL physics  before
recovering the canonical 2-D Fermi liquid regime only around 10~K (Figure \ref{Diag-phase}). 

Several experimental results show that the intermediate regime  10~K$ <
T < T_m$  behaves as a very unusual anisotropic 2-D metal.
As revealed by optical data in all Se-based salts, there still exists  at
$T = 20$~K a clear-cut gap in the frequency dependence of the
conductivity,
 $2\Delta_{\rho}\approx
200$ cm$^{-1}$ \cite{Vescoli98}.

This semiconducting-like spectrum contains 99\% of the
oscillator strength and the large DC conductivity is provided by a very
narrow zero frequency mode carrying the 1\% left over. As the spin
susceptibility remains finite in the same $T$-regime only 30-40\% below its
value at 300~K, the FIR gap in conductivity provides a remarkable
illustration for the spin-charge separation which still prevails
although the longitudinal DC transport looks like that of a Fermi liquid.
\begin{figure}[ htb] 
\epsfxsize0.95\hsize
\epsffile{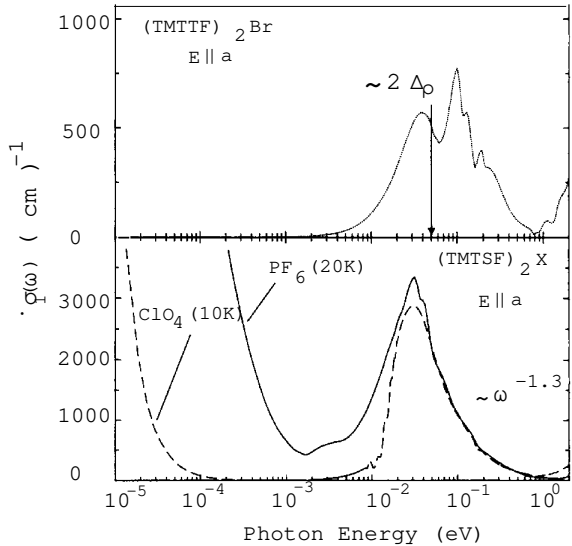}
\caption{One-chain optical conductivity of members of the (TM)$_2$X series in the normal phase at
low temperature. The arrow indicates location of the gap observed via the dielectric response in
the case of (TMTTF)$_2$Br, after \cite{Vescoli98}.  }    
\label{SigmaVescoli}   
\end{figure}    

As the temperature is decreased  below $T_m$, the anisotropy ratio $\rho_c/\rho_a$ remains 
temperature dependent, a feature that can be hardly tied with a Fermi liquid picture. It is only
below 10~K in (TMTSF)$_2$PF$_6$ under 9 kbar  that  the ratio becomes constant in temperature
suggesting a recovery of the usual Fermi liquid behavior (Figure \ref{anisotropy}).

	It is also interesting to notice that the existence of a $T_{m}$
marking the border between a high temperature Luttinger liquid and a
2D non-Fermi-liquid can be also extended to (TMTTF)$_{2}$PF$_{6}$
under pressure (Fig. \ref{Diag-phase}). Furthermore $T_{m}$ shows a very large
pressure coefficient. This property clearly rules out a simple relation
such as
$T_{m}\approx t_{\perp b}$  which is expected for the 2-D cross-over of
non-interacting electrons since 
$\delta\ln t_{\perp b}/\delta P\approx +2\% $ kbar$ ^{-1}$ from the
theory whereas $\delta\ln T_{m}/\delta P\approx +20\% $ kbar$ ^{-1}$
in
(TMTSF)$_{2}$PF$_{6}$ under ambient pressure. This feature suggests that
the beginning of the cross-over towards a 2D-NFL regime is renormalized
by the intra-chain interactions.

The power law coefficient can also be derived from the optical response
at frequencies
$\omega$ greater than the Hubbard gap since then the frequency dependent
conductivity is governed by inter Hubbard sub-band transitions and
becomes $\sigma (\omega) \approx
\omega^{4n^{2}K_{\rho}-5}$  (with $n = 2$ for the quarter-filled
situation). This is markedly different from the behavior of the
conductivity expected through the gap of a regular semiconductor where
$\sigma (\omega) \approx \omega^{-3}$. The experimental data of Figure~\ref{SigmaVescoli} for
(TMTSF)$_{2}$PF$_{6}$ and ClO$_4$ both give the exponent $- 1.3$ in the
power law frequency dependence in the low temperature range $(10-20$~K)
of the 2D intermediate phase. This exponent leads to $K_{\rho}=0.23$ 
in case of a quarter-filled 1-D band \cite{Giamarchi97}.

\section{Conclusion}

	This article aimed at surveying the physics of (TM)$_2X$ conductors in the high temperature regime,
{\it i.e.} in a regime too high in temperature for 3-D long range order (structural, magnetic or
superconducting) to become stable. 

We have shown a remarkable evolution throughout these Q-1-D conductors with commensurate
band-filling from Mott localized systems with a significant charge gap $2\Delta_{\rho} \approx$
1000 - 2000 cm$^{-1}$ in \sulf \  making the finite interchain kinetic coupling irrelevant at low
temperature all the way to the selenium-based conductors displaying a metallic-like behavior as far
as the longitudinal DC transport is concerned. However, a wealth of properties suggest that the
\sele\ 
 conducting salts cannot be viewed as canonical Fermi-liquid-like conductors. Transverse (along
$c^*$) and longitudinal resistivities exhibit opposite temperature dependencies above a
temperature marking the progressive establishment of a transverse coherence between chains in the
$(a-b)$ planes. Even below that cross-over temperature (which is $\approx  80$ K in
(TMTSF)$_2$PF$_6$ at ambient pressure)   remnants  of a non-Fermi behavior are the highlights of
these materials. A gap in the charge sector $( 2\Delta_{\rho} \approx $ 200~cm$^{-1}$) coexisting
with a DC conducting mode carrying only a minor fraction of the overall spectral weight and  far too
narrow to be understood in terms of a classical Drude model. These features suggest that the large
conductivity of
\sele\  salts developing at low temperature could possibly be ascribed to the existence of the
finite interchain coupling acting as an actual dopant in these otherwise commensurate Mott insulating
materials. In spite of some similarities, in the behavior of their transport properties, \sele\ 
and the anisotropic ruthenates, Sr$_2$RuO$_4$ \cite{Maeno94}, are likely to be very different since
it is believed that the latter conductor exhibits a coherent to incoherent transition along the
$c^{\star}$ axis around
100~K while the $c^{\star}$ transport   remains incoherent in (TMTSF)$_2$PF$_6$ down to low
temperature even though $\rho_{c^{\star}}$ behaves metal-like versus temperature.

Moving towards the low temperature regime, the concept of quasiparticle is recovered  {\it albeit}
with a severe renormalization as illustrated by the factor  $\approx 10$ enhancement measured in
the low energy spin sector via the nuclear spin-lattice relaxation rate of (TMTSF)$_2$ClO$_4$ at
ambient pressure or (TMTSF)$_2$PF$_6$ under pressure below 8~K \cite{Wzietek93}. It is the Fermi
nature of the 2-D electron gas which is responsible for the suppression of an itinerant
antiferromagnetic ground state in (TMTSF)$_2$PF$_6$
 under pressure. Furthermore, a mean-field theory based on a 2-D Fermi electron gas has also
been extremely successful in the low temperature domain explaining the restoration of a sequence of
ground states under a magnetic field exhibiting a non-commensurate magnetic modulation.
\cite{Gorkov84,Heritier84}.  The successful description of this cascade of field-induced phases in
terms of `BCS quasi-particles' is a bit puzzling, however, if one considers that the normal phase
under magnetic field  can be hardly tied in the Fermi liquid picture. The restoration of 
quasi-particles following the occurrence of long-range order is not specific to the Bechgaard
salts, it indeed finds a large echo in the context of high-$T_c$ cuprate superconductors for which
quasi-particles `\`a la BCS' are known to emerge in the  superconducting state in spite of the fact
that   the normal state  does not fit  at all with a Fermi liquid description
\cite{Shen95,Campuzano96}. Not less puzzling is a certain class  of   angular dependent
magnetotransport experiments performed at very low temperature in \sele \ 
\cite{Danner94,Naughton98}. These can be fittingly described by a classical Boltzmann theory which 
supports the recovery  of  coherent quasi-particles whose energy spectrum
 apparently shows no corrections due to many-body effects. This again occurs  in 
contradiction with  many anomalous properties  that show marginal part played by 
quasi-particles in a sizable part of the normal phase at high temperature.  The theoretical frame
that would solve this  paradox is not known so far.

The various ground states that are encountered in the (TM)$_2$X series have not been overviewed in
the present survey. To make a long story short let us mention that (TMTTF)$_2$PF$_6$ displays a
classical example for the spin-Peierls transition at $T _{\rm SP} = $18~K from a 1-D Mott insulator
to a spin singlet ground state accompanied by a lattice dimerization \cite{Pouget82}. The
spin-Peierls state turns into a commensurate N\'eel antiferromagnet under a pressure of about $8$
kbar (Fig.~\ref{Diag-phase}) \cite{Creuzet85,Caron88}. A recent detailed NMR under pressure study
displays a phase diagram in the vicinity of the critical pressure showing a competition between two
order parameters resulting in the suppression of the ordering temperature and the possible
existence of a quantum critical point
\cite{Chow98}. Increasing high pressure, the 1-D localization at high temperature is no longer
efficient as the cross-over temperature $T_m$ seen in $\rho_c$ rises under pressure from        
around 13 kbar. The N\'eel antiferromagnet turns into a SDW insulating state
\cite{Creuzet87,Klemme95}. Recent very high pressure data
\cite{Moser99} reveal a sharp suppression of the SDW phase transition above $40$ kbar and the
superconducting phase similar to what has been observed in (TMTSF)$_2$PF$_6$ at 9 kbar is expected
to become stable above 44 kbar or so.

The superconducting phase of 1-D organic conductors has not yet been intensively studied unlike the
superconducting phase of 2-D organic conductors. Even if theoretical predictions
\cite{Bourbonnais88} support the existence of an exotic pairing via the interchain exchange of
antiferromagnetic fluctuations experimental evidences are still scars. It has been known for a
long time that the sensitivity of the superconducting transition to defects is much higher than
what can be expected in a regular superconductor \cite{Bouffard82,Greene82}. A recent thermal
conductivity study of the superconducting state of (TMTSF)$_2$ClO$_4$ \cite{Belin97} has shown that
the superconducting order parameter cannot have nodes on the 1-D open Fermi surface. This finding
still leaves open various exotic scenarios: a spin-triplet pairing with the possibility of
reentrant superconductivity at very high magnetic fields aligned along the b' direction
\cite{Lebed98,Dupuis93,Lee97} or even d-wave pairing for
(TMTSF)$_2$ClO$_4$ in the presence of anion ordering at low temperature. There is still a lot more
work to be done on the superconducting phase of 1-D organic superconductors. The SDW ground state
of say,
(TMTSF)$_2$PF$_6$ has also been the subject of an intense effort
\cite{Jerome96,Gruner94, Takahashi86,Musfeldt95a,Musfeldt95c}. The whole condensate pinned by
impurities can be put into motion by an electric field larger than threshold fields of order 2~mV/cm
\cite{Tomic89} (instead of 100~mV/cm for CDW condensates) and gives rise to a collective conduction
channel. The systematic study of the SDW sliding conductivity
\cite{Kriza91} and the frequency dependence of the dielectric constant in a (TM)$_2$X alloy series
\cite{Traetteberg94} has shown the existence of two different channels of phason damping; one is due
to the dissipation of quasi-particles while the other one is related to the presence of defects.
Furthermore, the non-commensurate nature of the density wave goes hand in hand with its large
polarizability and a dielectric constant higher than 10$^{19}$ \cite{Mihaly91}.
 We are aware that this
article does not exhaust all properties of these fascinating materials, however we consider that
understanding in details their normal phase properties could also be of significant help for the
closely related systems such as the cuprate spin-ladder 1-D conductors which display superconductivity
 under pressure \cite{Uehara96}, although still in the presence of low lying spin excitations
\cite{Mayaffre98}. 

As we were finishing this article, H.J. Schulz passed away. It is the loss of a  colleague for the
whole physics community and for both of us in particular since Heinz has been a remarkable friend,
active and productive in the domain of low dimensional fermiology over the past 20 years.

\bibliography{articles,livres}

\begin{thebibliography}{100}

\bibitem{Hardy53}
G. Hardy and J. Hulm, Phys. Rev. {\bf 87},  884  (1953).

\bibitem{Matthias54}
B. Matthias, T. Geballe, S. Geller, and E. Corenwitz, Phys. Rev {\bf 95},  1435
   (1954).

\bibitem{Hulm80}
J. Hulm and B. Matthias, Science {\bf 208},  881  (1980).

\bibitem{Steglich79}
F. Steglich {\it et~al.}, Phys. Rev. Lett. {\bf 43},  1892  (1979).

\bibitem{Bednorz86}
J. Bednorz and K. Muller, Z. Phys. B {\bf 64},  189  (1986).

\bibitem{Little64}
W. Little, Phys. Rev. {\bf 134A},  1416  (1964).

\bibitem{Bardeen57}
J. Bardeen, L. Cooper, and J. Schrieffer, Phys. Rev. {\bf 108},  1175  (1957).

\bibitem{Akamatsu54}
H. Akamatsu, H. Inokuchi, and Y. Matsunaga, Nature {\bf 173},  168  (1954).

\bibitem{Coleman73}
L. Coleman {\it et~al.}, Solid State Comm. {\bf 12},  1125  (1973).

\bibitem{Ferraris73}
J. Ferraris, D. Cowan, V. Walatka, and J. Perlstein, J. Am. Chem. Soc. {\bf
  95},  948  (1973).

\bibitem{Schafer74}
D. Sch\"afer {\it et~al.}, Solid State Comm. {\bf 14},  347  (1974).

\bibitem{Groff74}
R. Groff, A. Suna, and R. Merrifield, Phys. Rev. Lett. {\bf 33},  418  (1974).

\bibitem{Denoyer75}
F. Denoyer, R. Com\`es, A. Garito, and A. Heeger, Phys. Rev. Lett. {\bf 35},
  445  (1975).

\bibitem{Peierls55}
R. Peierls, {\em Quantum Theory of Solids} (Oxford University Press, London,
  1955), p.108.

\bibitem{Horovitz75}
B. Horovitz, H. Gutfreund, and M. Weger, Phys. Rev. B {\bf 12},  3174  (1975).

\bibitem{Jerome82}
D. J\'erome and H. Schulz, Adv. in Physics {\bf 31},  299  (1982).

\bibitem{Friend78}
R.~H. Friend, M. Miljak, and D. J\'erome, Phys. Rev. Lett. {\bf 40},  1048
  (1978).

\bibitem{Megtert79}
S. Megtert {\it et~al.}, Solid State Commun. {\bf 31},  977  (1979).

\bibitem{Jacobsen78}
C. Jacobsen, K. Mortensen, J. Andersen, and K. Bechgaard, Phys. Rev. B {\bf
  18},  905  (1978).

\bibitem{Andrieux79}
A. Andrieux, C. Duroure, D. J\'erome, and K. Bechgaard, J. Phys. (Paris) Lett.
  {\bf 40},  381  (1979).

\bibitem{Pouget81}
J. Pouget, Chemica Scripta {\bf 55},  85  (1981).

\bibitem{Bechgaard80}
K. Bechgaard {\it et~al.}, Solid State Comm. {\bf 33},  1119  (1980).

\bibitem{Galigne79}
J. Galigne {\it et~al.}, Acta Cryst. {\bf B 35},  2609  (1979).

\bibitem{Brun77}
G. Brun {\it et~al.}, C.R. Acad. Sc. (Paris) {\bf 284 C},  211  (1977).

\bibitem{Coulon82}
C. Coulon {\it et~al.}, J. Phys. (Paris) {\bf 43},  1059  (1982).

\bibitem{Jerome91}
D. J\'erome, Science {\bf 252},  1509  (1991).

\bibitem{Haldane81}
F.~D.~M. Haldane, J. Phys. C {\bf 14},  2585  (1981).

\bibitem{Bourbon98}
C. Bourbonnais and D. J\'erome, Science {\bf 281},  1156  (1998).

\bibitem{Gorkov96}
L. Gor'kov, J. Phys. I (France) {\bf 6},  1697  (1996).

\bibitem{Jerome80}
D. J\'erome, A. Mazaud, M. Ribault, and K. Bechgaard, J. Phys. (Paris) Lett.
  {\bf 41},  L95  (1980).

\bibitem{Chaikin98}
D.~G. Clarke {\it et~al.}, Science {\bf 229},  2071  (1998).

\bibitem{Behnia95}
K. Behnia {\it et~al.}, Phys. Rev. Lett. {\bf 74},  5272  (1995).

\bibitem{Clarke97}
D.~G. Clarke and S. Strong, Adv. Phys. {\bf 46},  545  (1997).

\bibitem{Yakovenko98}
V. Yakovenko, preprint Cond-matter/9802172 (unpublished).

\bibitem{Naughton98}
I.~J. Lee and M.~J. Naughton, Phys. Rev. B {\bf 58},  R13 343  (1998).

\bibitem{Batail94}
V. Ilakovac {\it et~al.}, Phys. Rev. B {\bf 50},  7136  (1994).

\bibitem{Grant83}
P.~M. Grant, J. Phys. (Paris) Coll. {\bf 44},  847  (1983).

\bibitem{Yamaji82}
K. Yamaji, J. Phys. Soc. of Japan {\bf 51},  2787  (1982).

\bibitem{Ducasse86}
L. Ducasse {\it et~al.}, J. Phys. C {\bf 39},  3805  (1986).

\bibitem{Canadell94}
L. Balicas {\it et~al.}, J. Phys. I (France) {\bf 4},  1539  (1994).

\bibitem{Moret85}
R. Moret, J.~P. Pouget, R. Comes, and K. Bechgaard, J. Phys. (France) {\bf 46},
   1521  (1985).

\bibitem{Pouget96}
J.~P. Pouget and S. Ravy, J. Phys. I (France) {\bf 6},  1501  (1996).

\bibitem{Naughton88}
M.~J. Naughton {\it et~al.}, Phys. Rev. Lett. {\bf 61},  621  (1995).

\bibitem{Chaikin95}
S.~K. McKernan {\it et~al.}, Phys. Rev. Lett. {\bf 75},  1630  (1995).

\bibitem{Wang93}
W. Kang, S.~T. Hannahs, and P.~M. Chaikin, Phys. Rev. Lett. {\bf 70},  3091
  (1993).

\bibitem{Berthier98}
E. Hanson {\it et~al.}, proc. of ICSM'98 (unpublished).

\bibitem{Lebed96}
A.~G. Lebed, J. Phys. I (France) {\bf 6},  1819  (1996), and references
  therein.

\bibitem{Moret86}
R. Moret {\it et~al.}, Phys. Rev. Lett. {\bf 57},  1915  (1986).

\bibitem{Wudl83}
F. Wudl, D. Naleavek, J.~M. Torup, and N.~W. Extine, Science {\bf 22},  415
  (1983).

\bibitem{Emery83}
V.~J. Emery, J. Phys. (Paris) Coll. {\bf 44},  C3  (1983).

\bibitem{Bourbon84}
C. Bourbonnais {\it et~al.}, J. Phys. (Paris) Lett. {\bf 45},  L755  (1984).

\bibitem{Pines66}
D. Pines and P. Nozi\`eres, {\em The Theory of Quantum Liquids: Normal Fermi
  Liquids} (Benjamin, New York, 1966).

\bibitem{Miljak83}
N. Miljak, J.~R. Cooper, and K. Bechgaard, J. Phys. Coll. {\bf 44},  893
  (1983).

\bibitem{Miljak85}
N. Miljak and J.~R. Cooper, Mol. Cryst. Liq. Cryst. {\bf 119},  141  (1985).

\bibitem{Jacobsen86}
C.~S. Jacobsen, J. Phys. C {\bf 19},  5643  (1986).

\bibitem{Ducasse91}
A. Fritsch and L. Ducasse, J. Phys. I (France) {\bf 1},  855  (1991).

\bibitem{Ducasse96}
F. Castet, A. Fritsch, and L. Ducasse, J. Phys. I (France) {\bf 6},  583
  (1996).

\bibitem{Wzietek93}
P. Wzietek {\it et~al.}, J. Phys. I (France) {\bf 3},  171  (1993).

\bibitem{Mila95}
F. Mila and K. Penc, Phys. Rev. B {\bf 51},  1997  (1995).

\bibitem{Wilson75}
K.~G. Wilson, Rev. Mod. Phys. {\bf 47},  773  (1975).

\bibitem{Garoche82}
P. Garoche, R. Brusetti, and K. Bechgaard, Phys. Rev. Lett. {\bf 49},  1346
  (1982).

\bibitem{Brusetti83}
R. Brusetti, P. Garoche, and K. Bechgaard, J. Phys. C {\bf 16},  2495  (1983).

\bibitem{Pesty85}
F. Pesty, P. Garoche, and K. Bechgaard, Phys. Rev. Lett. {\bf 55},  2495
  (1985).

\bibitem{Gorkov95}
L.~P. Gor'kov, Europhys. Lett. {\bf 31},  49  (1995).

\bibitem{Moriya63}
J. Moriya, J. Phys. Soc. Jpn. {\bf 18},  516  (1963).

\bibitem{Soda77}
G. Soda {\it et~al.}, J. Phys. (Paris) {\bf 38},  931  (1977).

\bibitem{Bourbon89}
C. Bourbonnais {\it et~al.}, Phys. Rev. Lett. {\bf 62},  1532  (1989).

\bibitem{Bourbon93}
C. Bourbonnais, J. Phys. I (France) {\bf 3},  143  (1993).

\bibitem{Takahashi84}
T. Takahashi, D. J\'erome, and K. Bechgaard, J. Phys. (Paris) {\bf 45},  945
  (1984).

\bibitem{Takigawa86}
M. Takigawa and G. Saito, J. Phys. Soc. Jpn {\bf 55},  1233  (1986).

\bibitem{Devreux78}
F. Devreux and M. Nechtchein,  in {\em Proc. of the International Conference of
  Quasi-One-Dimensional Conductors I}, Edited by S. Barisic, A. Bjelis, J. R.
  Cooper and B. Leontic (Springer, Lectures Notes in Physics, Vol. 95, New
  York, 1978), p.\ 145.

\bibitem{Bourbon88b}
C. Bourbonnais {\it et~al.}, Europhys. Lett. {\bf 6},  177  (1988).

\bibitem{Azevedo82}
L.~J. Azevedo, J.~E. Schirber, and J.~C. Scott, Phys. Rev. Lett. {\bf 49},  826
   (1982).

\bibitem{Caretta95}
P. Caretta {\it et~al.},  in {\em Proc. of the Int. Conference on High Magnetic
  Field Phenomena-II Talahassee} (World Scientific, New Jersey, 1995), p.\ 328.

\bibitem{Gorkov98}
L.~P. Gorkov and M. Mochena, Phys. Rev. B {\bf 57},  6204  (1998).

\bibitem{Welber78}
B. Welber, P. Seiden, and P. Grant, Phys. Rev. B {\bf 18},  2692  (1987).

\bibitem{Megtert81}
S. Megtert {\it et~al.}, Solid State Comm. {\bf 37},  875  (1981).

\bibitem{Auban99}
P. Auban-Senzier (unpublished).

\bibitem{Gallois87}
B. Gallois, Ph.D. thesis, Univ. Bordeaux I, 1987.

\bibitem{Jacobsen81b}
C.~S. Jacobsen, K. Mortensen, M. Weger, and K. Bechgaard, Solid State Commun.
  {\bf 38},  423  (1981).

\bibitem{Moser98}
J. Moser {\it et~al.}, Eur. Phys. J. B {\bf 1},  39  (1998).

\bibitem{Fertey98}
P. Fertey, M. Poirier, and P. Batail, proc. of ICSM'98 (unpublished).

\bibitem{Jerome99}
L. Balicas and D. Jerome (unpublished).

\bibitem{Brown97}
S.~E. Brown {\it et~al.}, Synthetic Metals {\bf 86},  1937  (1997).

\bibitem{Laversanne84}
R. Laversanne {\it et~al.}, J. Phys. (Paris) Lett. {\bf 45},  L393  (1984).

\bibitem{Jacobsen81}
C.~S. Jacobsen, D. Tanner, and K. Bechgaard, Phys. Rev. Lett. {\bf 46},  1142
  (1981).

\bibitem{Jacobsen82}
C.~S. Jacobsen, D. Tanner, and K. Bechgaard, Mol. Cryst. Liq. Cryst. {\bf 79},
  261  (1982).

\bibitem{Jacobsen83}
C. Jacobsen, D. Tanner, and K. Bechgaard, Phys. Rev. B {\bf 28},  7019  (1983).

\bibitem{Kwak82}
J.~F. Kwak, Phys. Rev. B {\bf 26},  4789  (1982).

\bibitem{Yamaji90}
T. Ishiguro and K. Yamaji, {\em Organic Superconductors}, Vol.~88 of {\em
  Springer-Verlag Series in Solid-State Science} (Springer-Verlag, Berlin,
  Heidelberg, 1990).

\bibitem{Bourbon95}
C. Bourbonnais,  in {\em Les Houches, Session LVI (1991), Strongly interacting
  fermions and high-T$_c$ superconductivity}, edited by B. Doucot and J.
  Zinn-Justin (Elsevier Science, Amsterdam, 1995), p.\ 307.

\bibitem{Suzumura98}
Y. Suzumura, M. Tsuchiizu, and G. Gr\"uner, Phys. Rev. B {\bf 57},  R15 040
  (1998).

\bibitem{Vescoli98}
V. Vescoli {\it et~al.}, Science {\bf 281},  1181  (1998).

\bibitem{Timusk96}
N. Cao, T. Timusk, and K. Bechgaard, J. Phys. I (France) {\bf 6},  1719
  (1996).

\bibitem{Dressel96}
M. Dressel {\it et~al.}, Phys. Rev. Lett. {\bf 77},  398  (1996).

\bibitem{Giamarchi97}
T. Giamarchi, Physica {\bf B230-232},  975  (1997).

\bibitem{Mila96}
J. Favand and F. Mila, Phys. Rev. B {\bf 54},  10 425  (1996).

\bibitem{Pedron94}
D. Pedron, R. Bozio, M. Meneghetti, and C. Pecile, Phys. Rev. B {\bf 49},  10
  893  (1994).

\bibitem{Grioni97}
F. Zwick {\it et~al.}, Phys. Rev. Lett. {\bf 79},  3982  (1997).

\bibitem{Dardel93}
B. Dardel {\it et~al.}, Europhys. Lett. {\bf 24},  687  (1993).

\bibitem{Seidel83}
C. Seidel and V.~N. Prigodin, J. Phys. (Paris) Lett. {\bf 44},  L403  (1983).

\bibitem{Chaikin96}
P. Chaikin, J. Phys. I (France) {\bf 6},  1875  (1996).

\bibitem{Gorkov84}
L.~P. Gorkov and A.~G. Lebed, J. Phys. (Paris) Lett. {\bf 45},  L433  (1984).

\bibitem{Heritier84}
M. H\'eritier, G. Montambaux, and P. Lederer, J. Phys. (Paris) Lett. {\bf 45},
  L943  (1984).

\bibitem{Yamaji86}
K. Yamaji, Synthetic Metals {\bf 13},  19  (1986).

\bibitem{Chen86}
L. Chen, K. Maki, and V. Virosztek, Physica {\bf 143B},  444  (1986).

\bibitem{Dzyaloshinskii72}
I.~E. Dzyaloshinskii and A.~I. Larkin, Sov. Phys. JETP {\bf 34},  422  (1972).

\bibitem{Bychkov66}
Y.~A. Bychkov, L.~P. Gorkov, and I. Dzyaloshinskii, Sov. Phys. JETP {\bf 23},
  489  (1966).

\bibitem{Solyom79}
J. Solyom, Adv. Phys. {\bf 28},  201  (1979).

\bibitem{Barisic83}
S. Barisic, J. Phys. (Paris) {\bf 44},  185  (1983).

\bibitem{Barisic84}
S. Botric and S. Barisic, J. Phys. (Paris) {\bf 45},  185  (1984).

\bibitem{Barisic85}
S. Barisic, Mol. Cryst. Liq. Cryst. {\bf 119},  413  (1985).

\bibitem{Barisic81}
S. Barisic and S. Brazovskii,  in {\em Recent Developments in Condensed Matter
  Physics}, edited by J.~T. Devreese (Plenum, New York, 1981), Vol.~1, p.\ 327.

\bibitem{Mila94}
K. Penc and F. Mila, Phys. Rev. B {\bf 50},  11 429  (1994).

\bibitem{Emery82}
V.~J. Emery, B. R, and S. Barisic, Phys. Rev. Lett. {\bf 48},  1039  (1982).

\bibitem{Mila95b}
F. Mila, Phys. Rev. B {\bf 52},  4788  (1995).

\bibitem{Mila96b}
F. Mila and K. Penc (unpublished).

\bibitem{Bourbon91}
C. Bourbonnais and L.~G. Caron, Int. J. Mod. Phys. B {\bf 5},  1033  (1991).

\bibitem{Lieb68}
E. Lieb and F.~Y. Wu, Phys. Rev. Lett. {\bf 20},  1445  (1968).

\bibitem{Kimura75}
M. Kimura, Prog. Theor. Phys. {\bf 63},  955  (1975).

\bibitem{Gorkov73}
L.~P. Gor'kov and I.~E. Dzyaloshinskii, JETP Lett. {\bf 18},  401  (1973).

\bibitem{Brazo85}
S. Brazovskii and Y. Yakovenko, Sov. Phys. JETP {\bf 62},  1340  (1985).

\bibitem{Bourbon86}
C. Bourbonnais and L.~G. Caron, Physica {\bf 143B},  450  (1986).

\bibitem{Giamarchi91}
T. Giamarchi, Phys. Rev. B {\bf 44},  2905  (1991).

\bibitem{Shiba72}
H. Shiba, Phys. Rev. B {\bf 6},  930  (1972).

\bibitem{Maaroufi83}
A. Maaroufi {\it et~al.}, J. Phys. (Paris) Coll. {\bf 44},  1091  (1983).

\bibitem{Parkin83}
S. Parkin, J.~C. Scott, J.~B. Torrance, and E. Engler, J. Phys. (Paris) Coll.
  {\bf 44},  1111  (1983).

\bibitem{Voit95}
J. Voit, Rep. Prog. Phys. {\bf 58},  977  (1995).

\bibitem{Schulz90}
H.~J. Schulz, Phys. Rev. Lett. {\bf 64},  2831  (1990).

\bibitem{Mila93}
F. Mila and X. Zotos, Europhys. Lett. {\bf 24},  133  (1993).

\bibitem{Tutis90}
E. Tutis and S. Barisi\'c, Phys. Rev. B {\bf 42},  1015  (1990).

\bibitem{Pouget82}
J. Pouget {\it et~al.}, Mol. Cryst. Liq. Cryst. {\bf 79},  129  (1982).

\bibitem{Pouget97}
J.~P. Pouget and S. Ravy, Synthetic Metals {\bf 85},  1523  (1997).

\bibitem{Pouget76}
J.~P. Pouget {\it et~al.}, Phys. Rev. Lett. {\bf 37},  437  (1976).

\bibitem{Comes73}
R. Com\`es, M. Lambert, H. Launois, and H.~R. Zeller, Phys. Rev. B {\bf 8},
  571  (1973).

\bibitem{Schulz98}
H.~J. Schulz, G. Cuniberti, and P. Pieri, cond-mat/9807366 (unpublished).

\bibitem{LutherEmery74}
A. Luther and V.~J. Emery, Phys. Rev. Lett. {\bf 33},  589  (1974).

\bibitem{Emery79}
V.~J. Emery,  in {\em Highly Conducting One-Dimensional Solids}, edited by
  J.~T. Devreese, R.~E. Evrard, and V.~E. van Doren (Plenum Press, New York,
  1979), p.\ 247.

\bibitem{Gotschy92}
B. Gotschy {\it et~al.}, J. Phys. I (France) {\bf 2},  677  (1992).

\bibitem{Jerome94}
D. J\'erome,  in {\em Organic Conductors}, edited by J.~P. Farges (M. Dekker,
  New York, 1994), p.\ 405.

\bibitem{Klemme95}
B.~J. Klemme {\it et~al.}, Phys. Rev. Lett. {\bf 75},  2408  (1995).

\bibitem{Bourbon87}
C. Bourbonnais,  in {\em Low dimensional conductors and Superconductors},
  edited by D. J\'erome and L. Caron (Plenum, New York, 1987), p.\ 155, vol.
  155.

\bibitem{Creuzet87b}
F. Creuzet {\it et~al.}, Synthetic Metals {\bf 19},  277  (1987).

\bibitem{Azevedo81}
L.~J. Azevedo, J.~E. Schirber, R.~L. Greene, and E.~M. Engler, Physica {\bf
  B108},  183  (1981).

\bibitem{Schulz91}
H.~J. Schulz, Int. J. Mod. Phys. B  {\bf 5},  57  (1991).

\bibitem{Voit93}
J. Voit, Phys. Rev. B {\bf 47},  6740  (1993).

\bibitem{Schonhammer92}
V. Meden and K. Sch\"onhammer, Phys. Rev. B {\bf 46},  15753  (1992).

\bibitem{Degiorgi98}
V. Vescoli {\it et~al.}, Science {\bf 281},  1181  (1998).

\bibitem{Voit98}
J. Voit, Eur. Phys. J. B {\bf 5},  505  (1998).

\bibitem{Maeno94}
Y. Maeno {\it et~al.}, Nature {\bf 372},  532  (1994).

\bibitem{Shen95}
Z.~X. Shen and D.~S. Desseau, Phys. Rep. {\bf 253},  1  (1995).

\bibitem{Campuzano96}
J.~C. Campuzano {\it et~al.}, Phys. Rev. B {\bf 53},  R14 737  (1996).

\bibitem{Danner94}
G.~M. Danner, W. Kang, and P.~M. Chaikin, Phys. Rev. Lett. {\bf 72},  3714
  (1994).

\bibitem{Creuzet85}
F. Creuzet, D. J\'erome, and A. Moradpour, Mol. Cryst. Liq. Cryst. {\bf 119},
  297  (1985).

\bibitem{Caron88}
L.~G. Caron {\it et~al.}, Synthetic Metals {\bf 27B},  123  (1988).

\bibitem{Chow98}
D. Chow {\it et~al.}, Phys. Rev. Lett. {\bf 81},  3984  (1998).

\bibitem{Creuzet87}
F. Creuzet {\it et~al.}, Synthetic Metals {\bf 19},  289  (1987).

\bibitem{Moser99}
J. Moser (unpublished).

\bibitem{Bourbonnais88}
C. Bourbonnais and L. Caron, Europhys. Lett. {\bf 5},  209  (1988).

\bibitem{Bouffard82}
S. Bouffard {\it et~al.}, J. Phys. C {\bf 15},  295  (1982).

\bibitem{Greene82}
R. Greene {\it et~al.}, Mol. Cryst. Liq. Cryst. {\bf 79},  183  (1982).

\bibitem{Belin97}
S. Belin and K. Behnia, Phys. Rev. Lett. {\bf 79},  2125  (1997).

\bibitem{Lebed98}
A. Lebed and K. Yamaji, Phys. Rev. Lett. {\bf 80},  2697  (1998).

\bibitem{Dupuis93}
N. Dupuis, G. Montambaux, and C.~A. R.~S. de~Melo, Phys. Rev. Lett. {\bf 70},
  2613  (1993).

\bibitem{Lee97}
I.~J. Lee, M.~J. Naughton, G.~M. Danner, and P.~M. Chaikin, Phys. Rev. Lett.
  {\bf 78},  3555  (1997).

\bibitem{Jerome96}
D. J\'erome,  in {\em Physics and Chemistry of Low-Dimensional Inorganic
  Compounds}, edited by M.~G. C.~Schlenker, J.~Dumas and S. Smaalen (Plenum
  Press, New York, 1996), p.\ 141.

\bibitem{Gruner94}
G. Gr\"uner, Rev. Mod. Phys. {\bf 66},  1  (1994).

\bibitem{Takahashi86}
T.~T.~Y. Maniwa, H. Kawamura, and G. Saito, Physica {\bf 143B},  417  (1986).

\bibitem{Musfeldt95a}
J.~L. Musfeldt, M. Poirier, P. Batail, and C. Lenoir, Phys. Rev. B {\bf 51},
  8347  (1995).

\bibitem{Musfeldt95c}
J.~L. Musfeldt, M. Poirier, P. Batail, and C. Lenoir, Phys. Rev. B {\bf 52},
  15983  (1995).

\bibitem{Tomic89}
S. Tomic, J. Cooper, D. J\'erome, and K. Bechgaard, Phys. Rev. Lett. {\bf 62},
  462  (1989).

\bibitem{Kriza91}
G. Kriza {\it et~al.}, Phys. Rev. Lett. {\bf 66},  1922  (1991).

\bibitem{Traetteberg94}
O. Traetteberg {\it et~al.}, Phys. Rev. B {\bf 49},  409  (1994).

\bibitem{Mihaly91}
G. Mihaly, Y. Kim, and G. Gr\"uner, Phys. Rev. Lett. {\bf 66},  2806  (1991).

\bibitem{Uehara96}
M. Uehara {\it et~al.}, J. Phys. Soc. Jpn. {\bf 65},  2764  (1996).

\bibitem{Mayaffre98}
H. Mayaffre {\it et~al.}, Science {\bf 279},  345  (1998).

\end{thebibliography}
 
\end{document}